\documentclass[sigconf]{acmart}

\AtBeginDocument{%
  \providecommand\BibTeX{{%
    \normalfont B\kern-0.5em{\scshape i\kern-0.25em b}\kern-0.8em\TeX}}}

\setcopyright{acmcopyright}
\copyrightyear{2021}
\acmYear{2021}
\acmDOI{10.1145/1122445.1122456}

\acmConference[The Web Conference '22]{The Web Conference '22}{Apr 25th to 29th}{Lyon, France}
\acmBooktitle{The Web Conference '22, Apr 25th to 29th, Lyon, France}
\acmPrice{15.00}
\acmISBN{978-1-4503-XXXX-X/18/06}

\usepackage{graphicx}
\usepackage{algorithm}
\usepackage{algorithmic}
\usepackage{caption}
\usepackage{subcaption}
\usepackage{balance}  % for  \balance command ON LAST PAGE  (only there!)

\usepackage{multirow}

\usepackage{xcolor}
\usepackage{soul}

\title{What Is the Price of Data? A Measurement Study of Commercial Data Marketplaces}

\author{Santiago Andrés Azcoitia}
\email{santiago.azcoitia@imdea.org}
\affiliation{%
  \institution{IMDEA Networks Institute}
  \streetaddress{Av. Mar Mediterráneo, 22}
  \city{Leganes}
  \state{Madrid}
  \country{Spain}
}

\author{Costas Iordanou}
\email{kostas.iordanou@cut.ac.cy}
\affiliation{
  \institution{Cyprus University of Technology}
  \streetaddress{ Archiepiskopou Kyprianou 30, 3036}
  \city{Limassol}
  \country{Cyprus}
}

\author{Nikolaos Laoutaris}
\email{nikolaos.laoutaris@imdea.org}
\affiliation{%
  \institution{IMDEA Networks Institute}
  \streetaddress{Av. Mar Mediterráneo, 22}
  \city{Leganes}
  \state{Madrid}
  \country{Spain}
}

\begin{document}

%% Remove all unnecessary text related to the camera ready requirements of ACM
\setcopyright{none}
\settopmatter{printacmref=false, printccs=false, printfolios=false}
\renewcommand\footnotetextcopyrightpermission[1]{}
\pagestyle{plain}
\acmConference{}{}{}

\begin{abstract}
A large number of Data Marketplaces (DMs) have appeared in the last few years to help owners monetise their data, and data buyers fuel their marketing process, train their ML models, and perform other data-driven decision processes. In this paper, we present a first of its kind measurement study of the growing DM ecosystem and shed light on several totally unknown facts about it. For example, we show that the median price of live data products sold under a subscription model is around US\$1,400 per month. For one-off purchases of static data, the median price is around US\$2,200. We analyse the prices of different categories of data and show that products about telecommunications, manufacturing, automotive, and gaming command the highest prices. We also develop classifiers for comparing prices across different DMs as well as a regression analysis for revealing features that correlate with data product prices.
\end{abstract}


\maketitle

\section{Introduction}
\label{sect:introduction}
Data-driven decision making powered by ML algorithms is changing how the society and the economy work and is having a profound positive impact on our daily life. A McKinsey report predicted that data-driven decision-making could reach US\$2.5 trillion globally by 2025~\cite{McKinsey16}, whereas a recent market study within the scope of the European Data Strategy estimates a size of 827 billion euro for the EU27~\cite{EC20}. ML is driving up the demand for data in what has been called the fourth industrial revolution. Several companies, including Internet giants like Google, Facebook, and Amazon, already have access to most of the data required to train their ML algorithms. For the vast majority of companies though, big and small, and across sectors, there exist lots of data that they could exploit, but do not have or cannot collect. 

To satisfy this demand, several data marketplaces (hereinafter, DMs) have appeared in the last few years. DMs are mediation platforms that aim at connecting data providers (DPs, aka sellers) with potential buyers, and manage the transactions between them.  General-purpose DMs include AWS~\cite{AWS}, DAWEX~\cite{DAWEX}, DIH~\cite{DIH}, and Advaneo~\cite{Advaneo}. There also exist open data repositories, such as Dataverses~\cite{Dataverse} and Kaggle~\cite{Kaggle}, as well as specialised, or niche, DMs for specific industries, such as automotive~\cite{Otonomo,Caruso}, financial~\cite{Refinitiv, Battlefin}, marketing~\cite{LOTAME, LiveRamp}, and logistics~\cite{Veracity}, to name a few. 

An issue of paramount importance for DMs is that of \emph{data pricing}. Some DMs leave it to sellers to set a price for their data products, just like in any other marketplace for material goods or services. Many DMs do not list prices of their products, but leave it to buyers and sellers to agree on a price following a negotiation step. Like with material goods, pricing is a complex matter. In the case of DMs, however, it becomes even harder due to the  elusive nature of the traded “commodity”. Unlike oil, to which it is often compared~\cite{Humby06}, data can be copied / transmitted / processed with close to zero cost. This follows directly from its digital nature. Even the use of the term commodity is a gross oversimplification of what data is. Notice that whereas two litres of gasoline yield a similar mileage on two similar cars under similar driving styles, nothing of this sort applies to data since 1) two datasets of equal volume may be carrying vastly different amounts of usable information, 2) the same information may have tremendously different value for Service A than Service B, and 3) even if the per usage value of two services is the same, Service A may use the data 1,000 times more intensely than Service B leading to extremely different produced benefits. Some authors compared data to labor, too ~\cite{Arrieta18}. However, unlike labor, data is a non-rivalrous good meaning that its supply is not affected by its consumption, and thus selling data for a service A does not prevent a DP from selling (a copy of) the same data for a service B.

\begin{figure*}[t]
    \centering
    \includegraphics[width=\textwidth]{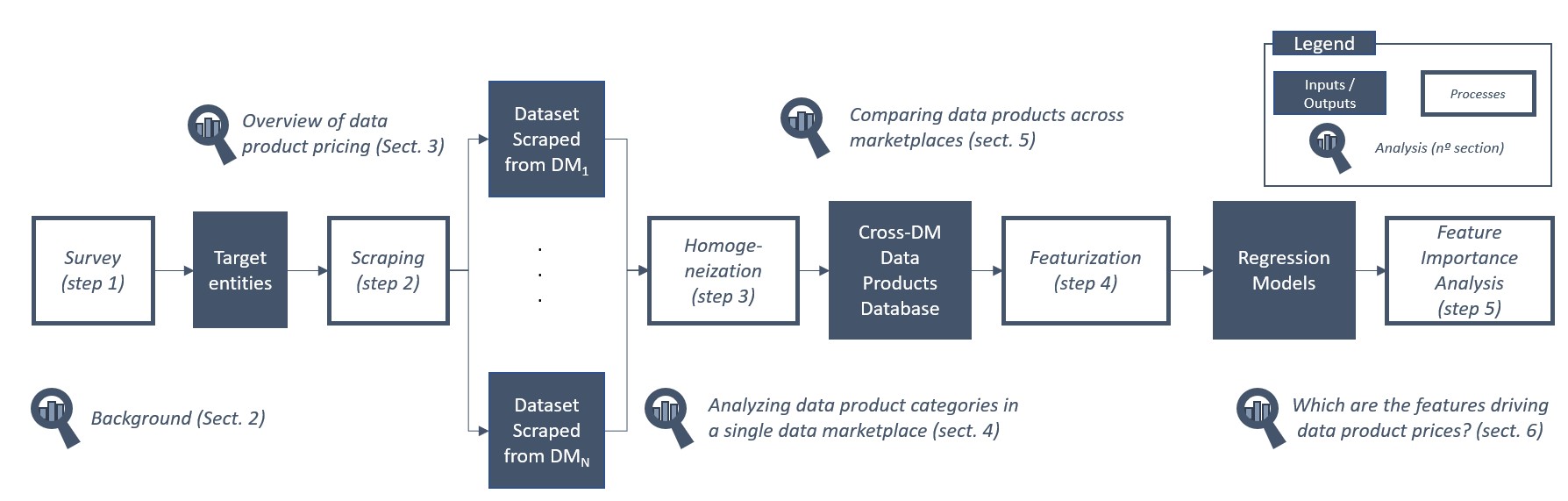}
    %\vspace{-8mm}
    \caption{Summary of our methodology}
    \label{FigMethodology}
\end{figure*}

The research community at the intersection between computer science and economics has studied several aspects of data pricing such as auction designs~\cite{Goldberg01, Goldberg03}, value-based pricing~\cite{Ohrimenko19, Agarwal19, Chen19}, pricing based on differential privacy~\cite{Ghosh11, Li15}, query pricing~\cite{Koutris12, Chawla19}, or quality-based pricing~\cite{Heckman15}. Still its elusive nature, and the complex business models under which it is made available makes it very hard to prescribe a price for data. Ultimately it is the market that decides and sets prices via complex mechanisms and feedback loops that are hard to capture. Despite the fact that there have been some works trying to measure the price of personal data of individuals~\cite{Carrascal13, Olejnik14, Papadopoulos17}, to the best of our knowledge, there is no systematic measurement study about the price of aggregated B2B data products traded in DMs like the ones mentioned above.   

\subsection{Our Contributions}
In this paper we present what is, to the best of our knowledge, the first systematic measurement study of DM for B2B data products.\footnote{We would like to clarify that companies like Axciom and Experian that have been the topic of public discussion due to data protection, targeted advertising and related matters are Data Providers (DP) in our terminology. Although we have included several of their products in the results of Sect.~\ref{sect:DataPricingOutlook} (they sell through AWS, DataRade, and other marketing \textit{niche} DMs), this paper is broader in scope and thus we do not pay special attention to them compared to any other of more than 2,000 sellers of our study.} This ecosystem, despite being quite vibrant commercially, remains completely unknown to the scientific community. Very basic questions such as ``\emph{What is the range of prices of data traded in modern DMs?}'', ``\emph{Which categories and types of data products command the highest prices?}'', ``\emph{Are the observed prices in one DM consistent with the prices for similar products in other DMs?}'', ``\emph{Which are the features, if any, that correlate with the most expensive data products?}'' appear to have no answer and evade most meaningful speculations.

To answer such questions we first conducted an extensive survey for compiling a catalogue with more than 180 DMs. We then selected 9 of them that fulfill necessary criteria for a measurement study. For these DMs we developed custom crawlers for retrieving information about the products they trade. Using these crawlers, and adding the portfolio of another 29 DPs, we obtained information for more than 210,000 data products and a catalog of more than 2,000 distinct sellers. We also developed machine learning classifiers for identifying data products of similar categories to compare prices across DMs, and executed 9 different regression models to understand which features are driving the prices of data products.

%\vspace{-2mm}
\subsection{Our Findings}
Analysing the collected data we observed that the majority of data products were either given for free, or did not carry a fixed price, but rather were up for direct negotiation between the seller and interested buyers. Focusing on the ones that carried a price, some 4,200 of them, we observed the following: 

\begin{itemize}
    \item Prices vary in a wide range from a few US-dollars up to several hundred of thousands. The median price for data products sold under a \emph{subscription} model is US\$1,400 per month. The median price for products sold as an \emph{one-off} purchase is US\$2,200.
    \item Focusing on Amazon Web Services' (AWS) DM we analysed products of different categories and found that those related to \emph{telecoms}, \emph{manufacturing}, \emph{automotive} and \emph{gaming} command the highest median prices.
    \item Using our classifiers, we enriched our sample by consistently labelling data products according to AWS's defined categories.
    \item Using regression models, we managed to fit the prices of commercial products from their features with accuracy ($R^2$ score) above 0.84.
    \item Features related to volume and units, and domain-specific characteristics of data products captured by their descriptions and categories are responsible for 66\% of this score.
    \item Due to the heterogeneity of the sample there is no single feature other than volume units (with exceptions) that drives the prices, but instead we spotted meaningful features that proved to be conclusive in specific domains: stems like `\textit{custom}', `\textit{edgar}' or `\textit{market}', and `\textit{contact}', `\textit{identifi}' or `\textit{accur}' appear in the top 10 for \emph{financial} and \emph{marketing} products, respectively. Interestingly, data update rate seem to be a key price driver for \emph{financial} and \emph{healthcare}-related products, whereas the ability to provide exact locations and the possibility of reconstructing sessions (i.e., connecting individual data points from the same owner) are for \emph{marketing} data.
    \item We also studied temporal aspects of DMs and noticed that DMs such as AWS have been growing with a significant 3\% monthly rate from December 2020 to August 2021.
\end{itemize}

Like in all measurement studies of Internet-scale phenomena, we'll refrain from claiming that any of our findings are ``typical'' or ``representative''. What we do claim, however, is that to the best of our knowledge, our measurement study is the first one that attempts to characterise the DM sector, and our above mentioned quantitative results were previously totally unknown. Also, as it will become evident from our methodology later, and to the best of our knowledge, we collected all publicly available DM pricing information that was accessible during the time of our study.   

The remainder of the paper is structured as shown in Fig.~\ref{FigMethodology}. First, we introduce the data trading ecosystem to frame the scope of our analysis and show some initial outcomes of our measurement study in Sect.~\ref{sect:DataTradingEcosystem}. In Sect.~\ref{sect:DataPricingOutlook}, we present a novel analysis on data product pricing in commercial DMs. Furthermore, Sect.~\ref{sect:Analyzing productsInAsingleDM} dives deeper into analyzing AWS' DM, which accounts for the highest number of price references in our sample. We then develop tools for enriching our sample and allowing cross-DM price comparison in Sect.~\ref{sect:CrossDMcomparison}. Finally, in Sect.~\ref{sect:UnderstandingFeaturesDrivingDataPrices}, we apply several methodologies for analysing the importance of different metadata features in determining the price of commercial data products.

\section{Background} 
\label{sect:DataTradingEcosystem}

\subsection{Terminology}
\textit{Data Providers} (DP) or sellers are entities that provide \textit{data} as a product, be they raw, enriched data, access to information through a GUI, or information contained in reports. They may combine data from different sources to enrich their products and increase its value. Examples of DPs are BookYourData~\cite{BookYourData}, Benzinga~\cite{Benzinga}, and Enigma~\cite{Enigma}. 

\textit{Data marketplaces} (DM) are two-sided mediation platforms aimed at liaising data providers with potential buyers, and managing data transactions between them. Such transactions oftentimes involve some kind of economic exchange, which is also controlled by the platform. Examples of DM are AWS marketplace~\cite{AWS}, Dawex~\cite{DAWEX}, Battlefin~\cite{Battlefin}, or DataRade~\cite{DataRade}.

\textit{Personal Information Management Systems} (PIMS) are platforms aimed at empowering individuals to take control of their personal information (PI). PIMS leverage recent data protection laws to let users manage their PI stored by Internet service providers, and their consent for such data to be shared with third parties for specific purposes. Some of them offer options for users to monetise their PI, as well. For example, PIMS like Wibson~\cite{Wibson}, or digi.me~\cite{digi.me} allow a buyer to acquire PI of their users. In the rest of the paper we will use the term DM to refer to any kind of data intermediary, be it a PIMS or a B2B DM.

\begin{figure*}[t]
    \centering
    \begin{subfigure}[b]{0.20\textwidth}
        \centering
        \includegraphics[width=\textwidth]{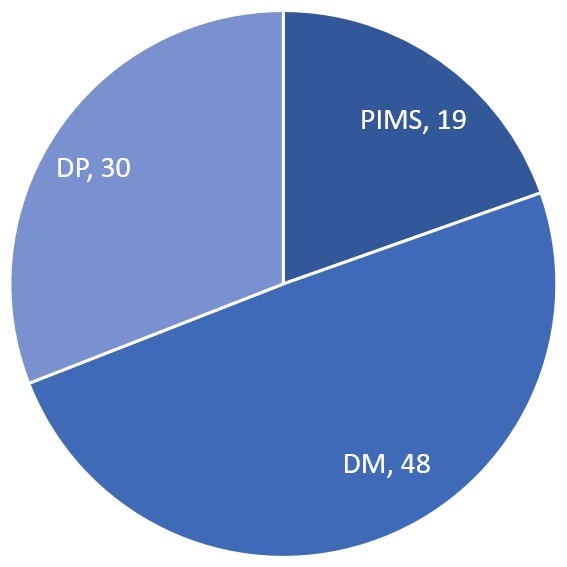}
        \caption{Nº entities by business model}
        \label{SubFigEntityByBSSModel}
    \end{subfigure}
    \begin{subfigure}[b]{0.38\textwidth}
        \centering
        \includegraphics[width=\textwidth]{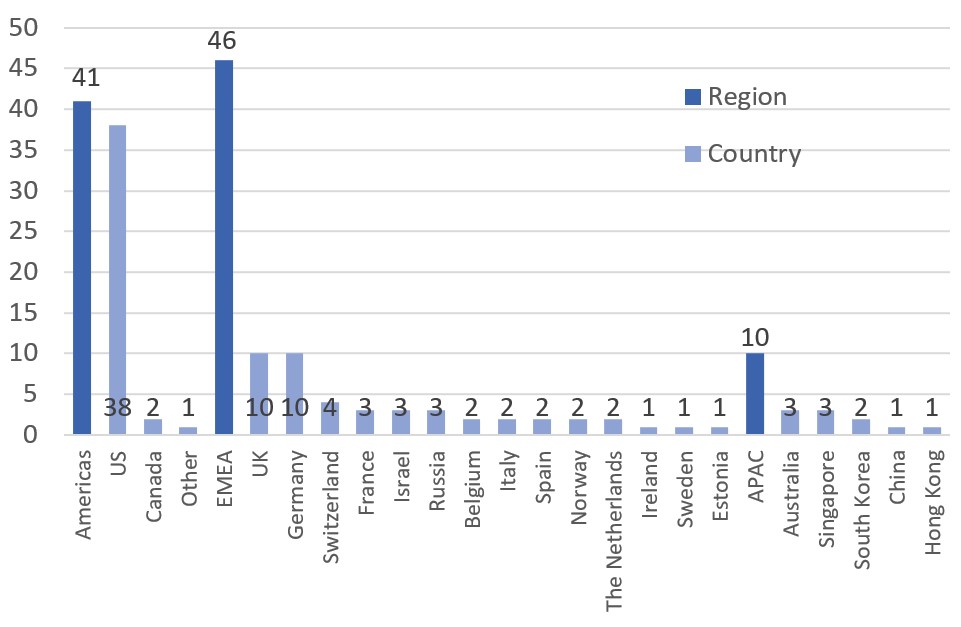}
        \caption{Nº entities by Country}
        \label{SubFigEntitiesByCountry}
    \end{subfigure}
    \begin{subfigure}[b]{0.38\textwidth}
        \centering
        \includegraphics[width=\textwidth]{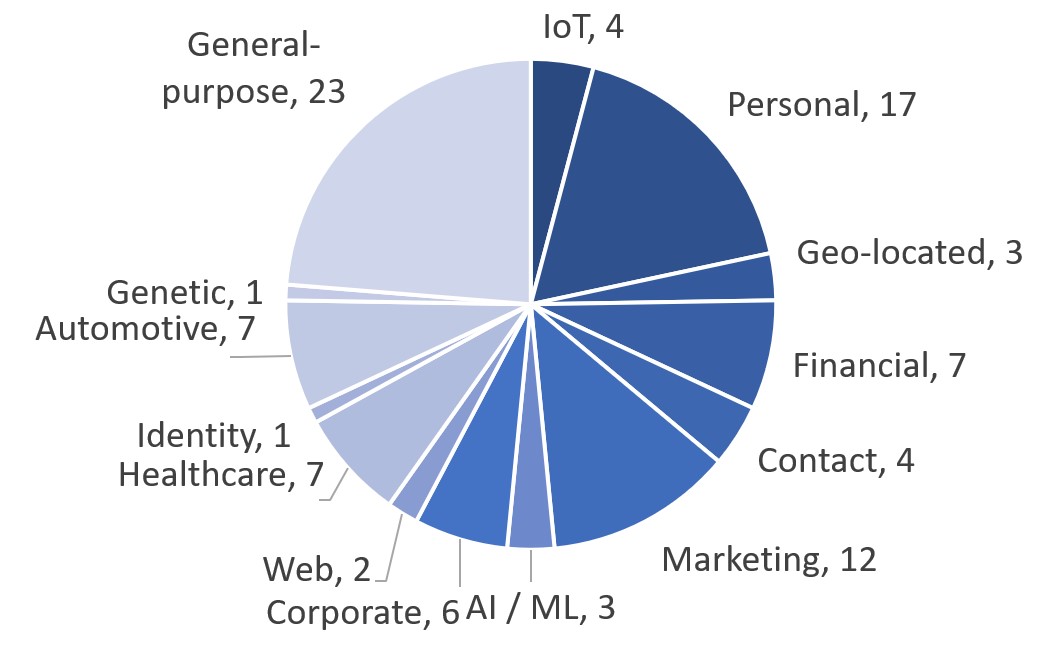}
        \caption{Nº entities by data type}
        \label{SubFigEntitiesByDataType}
    \end{subfigure}
    %\vspace{-3mm}
    \caption{Scope of our survey on entities trading with data}
    \label{FigSurveyScope}
\end{figure*}

Data delivery methods usually depend on the product and determine the pricing mechanism. Traditionally DMs offered bulk downloads of datasets in one or more files once the buyer satisfied a one-off payment defined by the seller, or requested a direct contact to understand the buyer's needs and deliver a tailored product at a personalised price. With the advent of cloud computing, digitalization and IoT, data is increasingly demanded in real time. Such real-time data is usually delivered through APIs or data feeds and charged by volume or usage. In addition, delivery through temporary URLs, or revocable PK to decrypt data feeds is gaining \textit{momentum}, and buyers usually charge sellers based on time subscriptions in such settings. Finally, some DMs offer access to data via web-services that allow buyers to make queries and export results in different formats.

\subsection{Bird's eye view of the DM landscape}

\begin{figure}[t]
    \centering
    \includegraphics[width=0.45\textwidth]{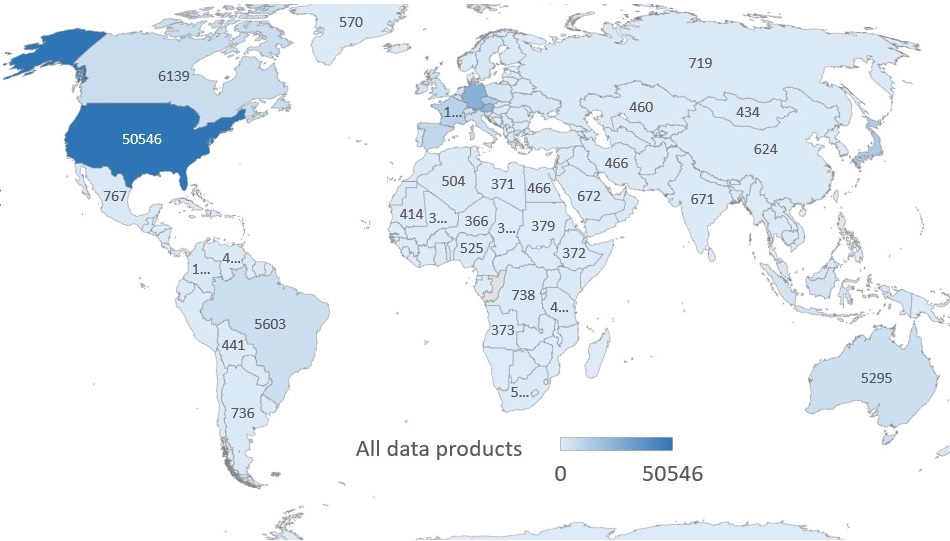}
    %\vspace{-3mm}
    \caption{Data products by country}
    \label{fig:DataProductsByCountry}
\end{figure}

We conducted an extensive web-search, as well as consulted with experts in the area to compile a list of existing DMs. This resulted in identifying around 180 DMs. The `\textit{Marketplaces for Data: An Initial Survey}' ~\cite{Schomm13} and `\textit{Data Marketplaces: Trends and Monetisation of Data Goods}' papers of 2012 and 2019 mention 46 and 16 such DMs and, therefore, we believe that we have covered a good part of the DM market. Out of those 180, we selected 97 of them for a more in-depth study. We discarded concept projects, online advertising platforms, and Internet service providers not offering data products. Moreover, we balanced our selection in terms of the business models covered, and included entities from 17 countries that trade very different types of data as Fig.~\ref{FigSurveyScope} shows.

The 97 selected entities were analysed from a business perspective for a related market report in terms of their exact business models, their data delivery mechanisms, and so on. The details of this report are not relevant to this paper and therefore we will only present a high-level summary. In Fig.~\ref{SubFigEntityByBSSModel} we present a break down in terms of their exact type as defined before. We see that more than 40\% are DMs%, i.e., inter-mediation platforms between data buyers and sellers
. Figure~\ref{SubFigEntitiesByCountry} depicts their geographic spread, indicating that most of them are in the US and Europe. Finally, and most interestingly, in Fig.~\ref{SubFigEntitiesByDataType} we present a breakdown \emph{by category} of the data made available by these 97 DMs. 23 out of them are general-purpose DM trading any type of data. We will dive deeper into the category of data products sold by the largest of these DMs in Sect.~\ref{sect:Analyzing productsInAsingleDM}. Obtaining this last result across DMs is more challenging, since they use different categorisation systems and criteria to assign such categories to data products. Later in Sect.~\ref{sect:CrossDMcomparison} we will explain how we homogenised the categories across the different DMs. For now we will only note that personal information of individuals (anonymised and, often, aggregated) is the most popular category of data among specialized DMs, followed by marketing and corporate data.

%\vspace{-3mm}
\subsection{Compiling a dataset of DM data products}

From our analysis of the aforementioned DMs we identified a subset that fulfilled a number of criteria for using them as sources of data for a measurement study. Such criteria include that they grant access to their product catalog without requiring an account, or through an account but without a vetting process or upfront paid registration, that they have a reasonably large catalog that includes sufficient descriptions of their data products, and that they include a clear description of their pricing policy. Out of the 180 initial DMs, only 9 companies fulfilled all of the above criteria. Most of the DMs did not make it to the list simply because they do not allow non-paying users to browse their catalogs. For example, marketing-related private marketplaces such as \textit{Liveramp}, \textit{LOTAME} or \textit{TheTradeDesk} neither provide public per-product information nor any price references. However, they do provide information about their data partners, which we included in our catalog, and we did find that 45\% of DPs in those DMs sell through general-purpose DMs such as \textit{AWS} or \textit{DataRade}, as well. We also discarded several otherwise \textit{scrapable} general-purpose DM such as \textit{Data Intelligence Hub} (DIH), \textit{Google Cloud DM} because they included only free data products. From these free open data marketplaces, we chose to scrape the largest one, \textit{Advaneo}, to help in training data product category classifiers.

Table~\ref{tab:marketplaces} lists the 9 DMs that were used as sources of data for our study. They include 5 general-purpose DMs, and 4 niche DMs.  In addition, we included 29 DPs commercializing their own 776 data products without intermediaries. We developed our own web crawler to render and download web pages from them, and specialised parsers for extracting metadata from each one. We followed common crawling good practices ~\cite{Hils20}. For example, we avoided visiting several times the same product page in each scraping round and we set up a random wait time from 1 to 2 minutes after requesting a web page in order to avoid flooding the target servers with requests. 

We collected information about 213,964 data products from 1,853 distinct sellers in total (2,015 including  partners of private marketing DMs). We scraped all available metadata such as the product id, title and description, source, seller and, when available, its geographic scope, volume, category, use cases, history available, format, etc. We searched for and eliminated duplicates from a single seller within the same DM. We paid special attention to information related to pricing, such as its model, options and any details about the actual prices of data products.  

Regarding the geographical scope of data products, we found that DMs aggregate information from different countries. 14,472 (7\%) of the products did not inform about their scope, and 1,177 (11\% out of the 10,772 paid products) claimed to be global (e.g., those relying on a worldwide accessible mobile app to generate the data). Figure~\ref{fig:DataProductsByCountry} shows the number of data products covering each country. Regarding the number of \emph{paid} data products, US leads this ranking: around 30\% of paid products cover this country. Canada (9.3\%), UK (9.2\%), Germany (7.6\%), France (7.4\%), and Spain (7.1\%) follow the US in the ranking of countries by number of \emph{paid} products.

\section{Overview of data product pricing} 
\label{sect:DataPricingOutlook}

It may appear initially surprising that, despite being commercial entities in the B2B space, most of the surveyed and some of the scraped DMs, offer predominately free (most of the time open) data. Again we point to the fact that these are privately held companies (e.g., Advaneo~\cite{Advaneo} and DIH~\cite{DIH}) and not open data NGOs or government initiatives. Our conjecture is that since DMs are two-sided platforms, pre-populating them with free data is a very reasonable bootstrapping strategy, since it can attract the initial ``buyers'', which in turn will attract commercial sellers and thus help the DM grow its revenue.  

%Surprisingly, most of the products being offered by some commercial DM can be downloaded for free, are often unclassified and contain unprocessed open data. In particular, Advaneo~\cite{Advaneo}, DIH~\cite{DIH} were found to be basically open data aggregators, whereas Google Cloud DM ~\cite{GCM} offers samples and references to sellers in addition.

\begin{table}[t]
\caption{Summary of scraped DMs} 
%\vspace{-4mm}
\label{tab:marketplaces}
\begin{center}
\begin{tiny}
\resizebox{\columnwidth}{!}{%
\begin{tabular}{l|c|c|c}
\toprule
\hline
\textbf{Marketplace}&\#\textbf{Data Products}&\#\textbf{Paid products}&\#\textbf{Sellers}\\
%\midrule
\hline
\textbf{Advaneo}&198,743&1&N/A\\ \hline
\textbf{AWS}&4,013&2,515&262\\ \hline
\textbf{DataRade}&1,592&1,592&1,262\\ \hline
\textbf{Knoema}&158&158&142\\ \hline
\textbf{DAWEX}&160&160&79\\ \hline
\textbf{Carto}&8,182&5,283&42\\ \hline
\textbf{Crunchbase}&16&14&15\\ \hline
\textbf{Veracity}&115&95&38\\ \hline
\textbf{Refinitiv}&214&185&76\\ \hline
%\textbf{Liveramp}&N/A&N/A&135\\ \hline
%\textbf{LOTAME}&16&16&128\\ \hline
%\textbf{The Trade Desk}&N/A&N/A&122\\ \hline
\textbf{Other data providers}&771&769&29\\ \hline
\bottomrule
\end{tabular}
}
\end{tiny}
\end{center}
\end{table}

In this section we will focus on the 10,772 paid data products, for which we managed to extract some information about their pricing. Despite being few compared to the free ones, this sample provides very valuable insights about the current status of commercial DMs, as well as to where this segment of the economy is heading to, and how.  

There is a great magnitude of pricing schemes for data products, such as seller-led, buyer-led (bidding), revenue-sharing, tiered-pricing, negotiation-based, usage-based, etc.~\cite{Pei20, Loser13}. Predominant among the 10,772 non-free data products are the \emph{subscription-based} model, and the \emph{one-off} model, seller-led in both cases. The first one is used mostly for ``live'' data usually accessed via an API (e.g., IoT data), whereas the second is used with static siloed data, usually downloaded as one or more files.

\begin{figure}[t]
    \centering
    \begin{subfigure}[b]{0.45\textwidth}
        \centering
        \includegraphics[width=\textwidth]{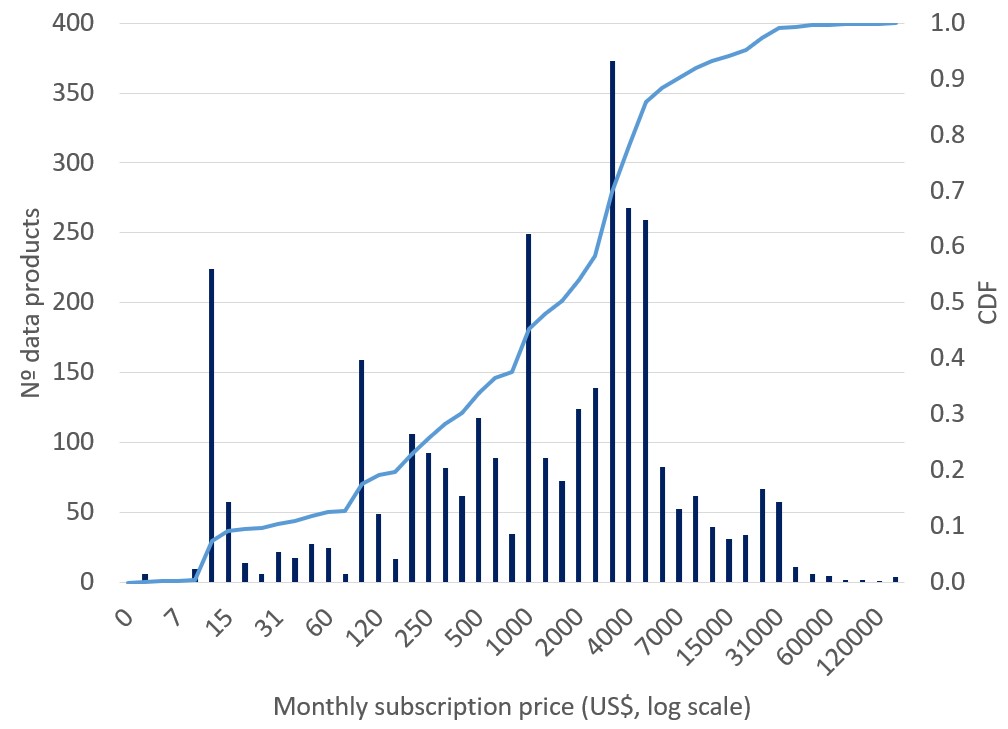}
        %\vspace{-5mm}
        \caption{Subscription-based}
        \label{SubFigHistogramSubscriptionBased}
    \end{subfigure}
    \begin{subfigure}[b]{0.45\textwidth}
        \centering
        \includegraphics[width=\textwidth]{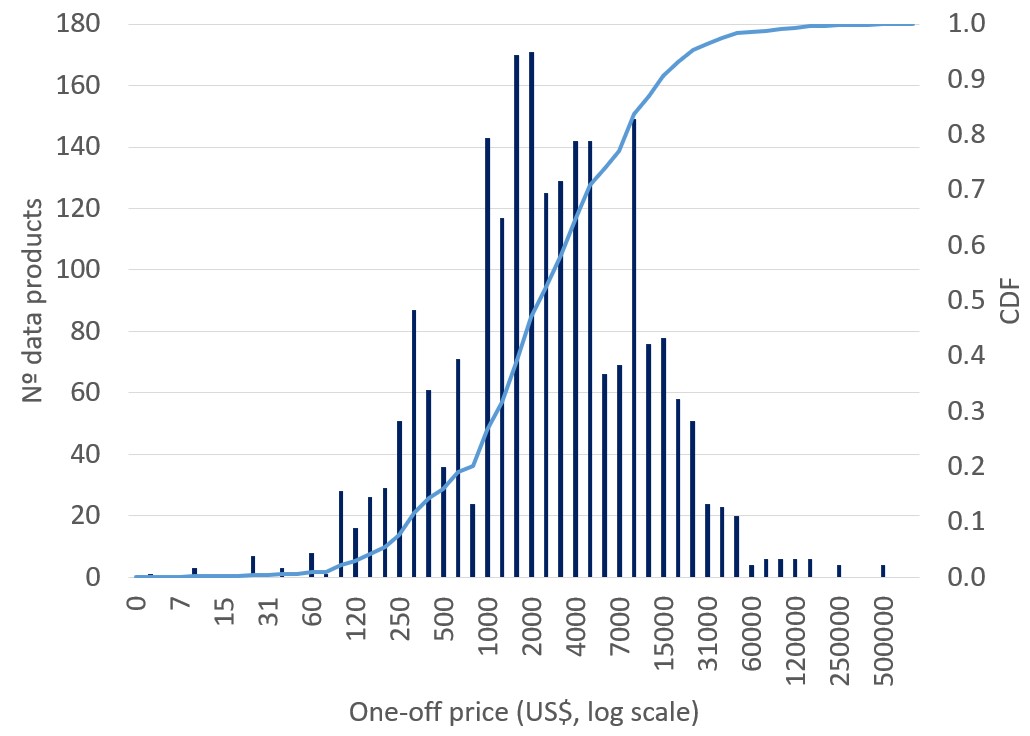}
        %\vspace{-5mm}
        \caption{Fixed price (one-off)}
        \label{SubFigHistogramOneOff}
    \end{subfigure}
    %\vspace{-3mm}
    \caption{Histogram and CDF of data products}
    \label{FigPriceHistograms}
\end{figure}

4,162 products from 443 distinct providers provided clear information about their prices. Figure~\ref{SubFigHistogramSubscriptionBased} shows a histogram and the corresponding CDF of monthly prices for data products offered under a \emph{subscription model}. 

We see prices across a wide range up to US\$150,000 per month. Doing some manual analysis we realised that cheap products costing less than US\$100 per month, are often curated and cleaner versions of data from open sources. For example, a seller offers a historical compilation of quarterly reports submitted to the US Securities and Exchange Commission (SEC), which can be downloaded from their websites, as well. They also include several low-cost ``promotion samples'' of more expensive products from well-known sellers, such as GIS data and supporting metadata for a small area of some US cities. The median price is US\$1,417 per month. Almost one-third of all data products, including targeted market data and reports for example, are sold for US\$2,000-5,000 per month.

Figure~\ref{SubFigHistogramOneOff} depicts the prices for data products sold under a \emph{one-off model}. Comparing this figure with the previous one for subscription-based access we quickly notice that: (1) one-off data products tend to be more expensive: median price US\$2,176 vs. US\$1,417 per month for subscription-based products; maximum price US\$500,000, more than 3 times higher than the maximum in subscription-based access, and (2) one-off products have a price histogram with fewer modes that is more normally distributed around its median at US\$2,176. Despite the heterogeneous set of products within the US\$1,000-4,000 interval, we found that voluminous targeted contact data products are contributing to the maximum. 

Most interestingly, we observe a long tail of valuable data products in both Fig.~\ref{SubFigHistogramSubscriptionBased} and ~\ref{SubFigHistogramOneOff}. We will come back to these products in a later section.

%\vspace{-5mm}
\section{Analyzing data product categories in a single data marketplace}
\label{sect:Analyzing productsInAsingleDM}
In order to get a more in-depth understanding of data pricing we analysed the catalog of the Amazon Web Service (AWS) DM, the one with the largest base of paid products with prices among the ones presented in Tab.~\ref{tab:marketplaces}.

AWS classifies data products by \textit{category}. Specifically, a product can belong to none, one, or several categories corresponding to industries or sectors of the economy. For instance, credit cards transaction data products are classified both as `\textit{Financial}' and `\textit{Retail, Location and Marketing}', whereas weather related ones are not labelled in any category. We mark such unclassified products as `\textit{Other}'. 

\begin{figure}[t]
    \centering
    \includegraphics[width=0.45\textwidth]{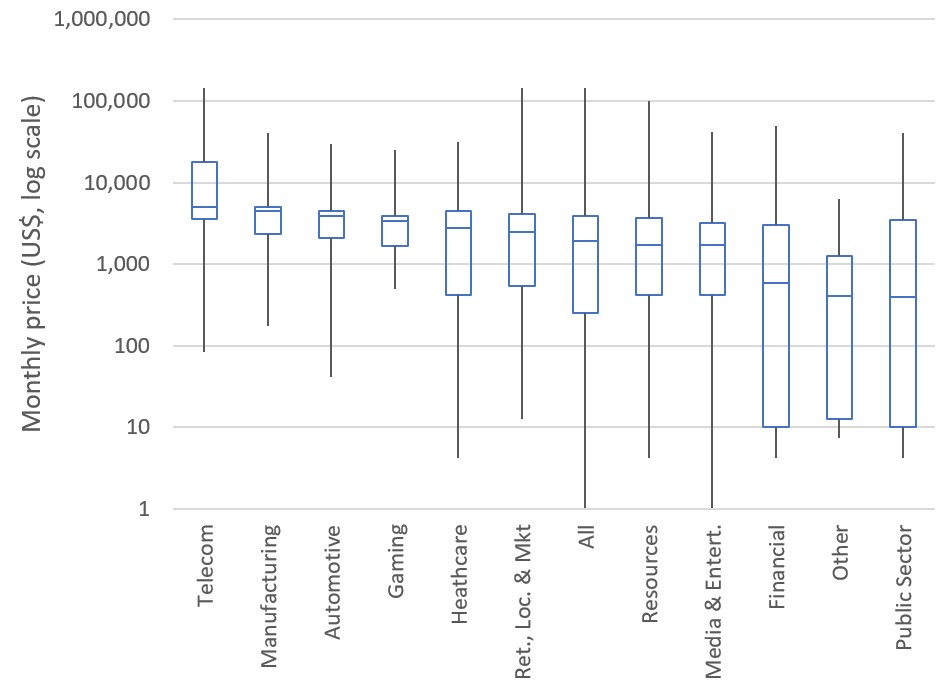}
    %\vspace{-3mm}
    \caption{Box plot of subscription-based data prices by industry in AWS. M\&E stands for media and entertainment. \textit{Retail} includes products related to \textit{Retail, Location and Marketing}. \textit{Other} means the product does not belong to any category.}
    \label{FigAWSCDFbyIndustry}
\end{figure}

Figure~\ref{FigAWSCDFbyIndustry} shows a box plot of data products by category in AWS. The X-axis shows the different categories ordered in decreasing median price, whereas the Y-axis represents the monthly price to get access to the data. `\textit{Telecom}', `\textit{Manufacturing}' and `\textit{Automotive}' categories exhibit a median price significantly above the global ($\times2.6$, $\times2.3$ and $\times2$, respectively). Most low-value products belong to the \textit{Public Sector}, \textit{Financial} (stock price feeds for example), and \textit{Other} categories.

\begin{figure}[t]
    \centering
    \begin{subfigure}[b]{0.45\textwidth}
        \centering
        \includegraphics[width=\textwidth]{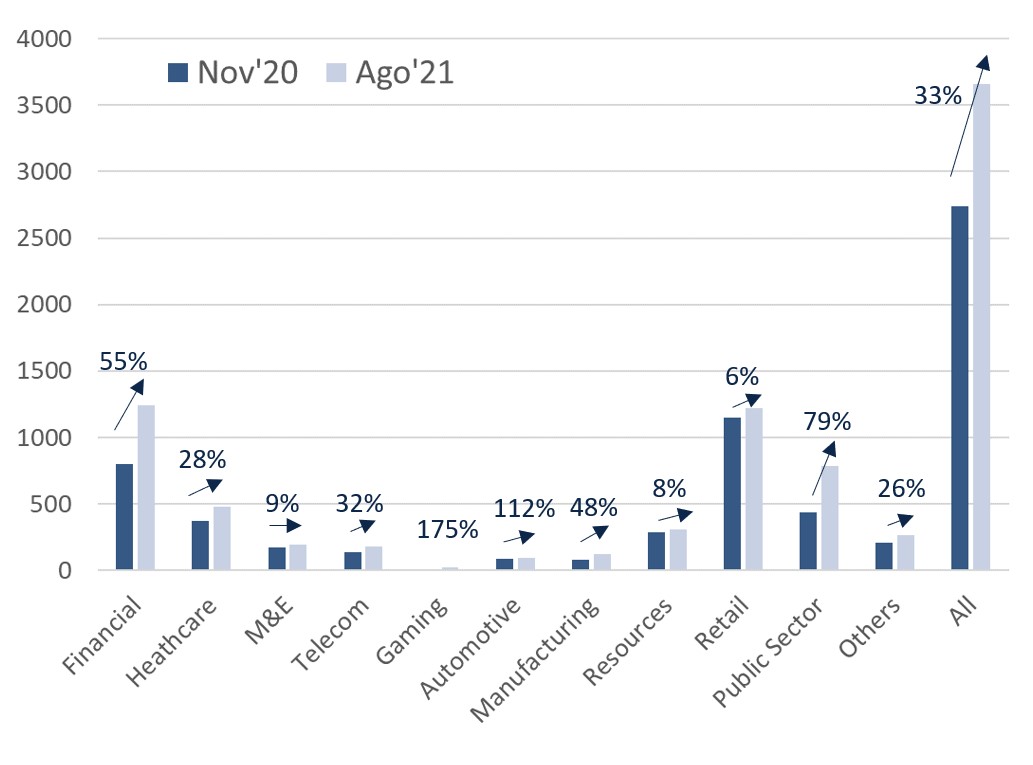}
        %\vspace{-5mm}
        \caption{Nº products by category}
        \label{SubFigAWSProductEvolution}
    \end{subfigure}
    \hfill
    \begin{subfigure}[b]{0.45\textwidth}
        \centering
        \includegraphics[width=\textwidth]{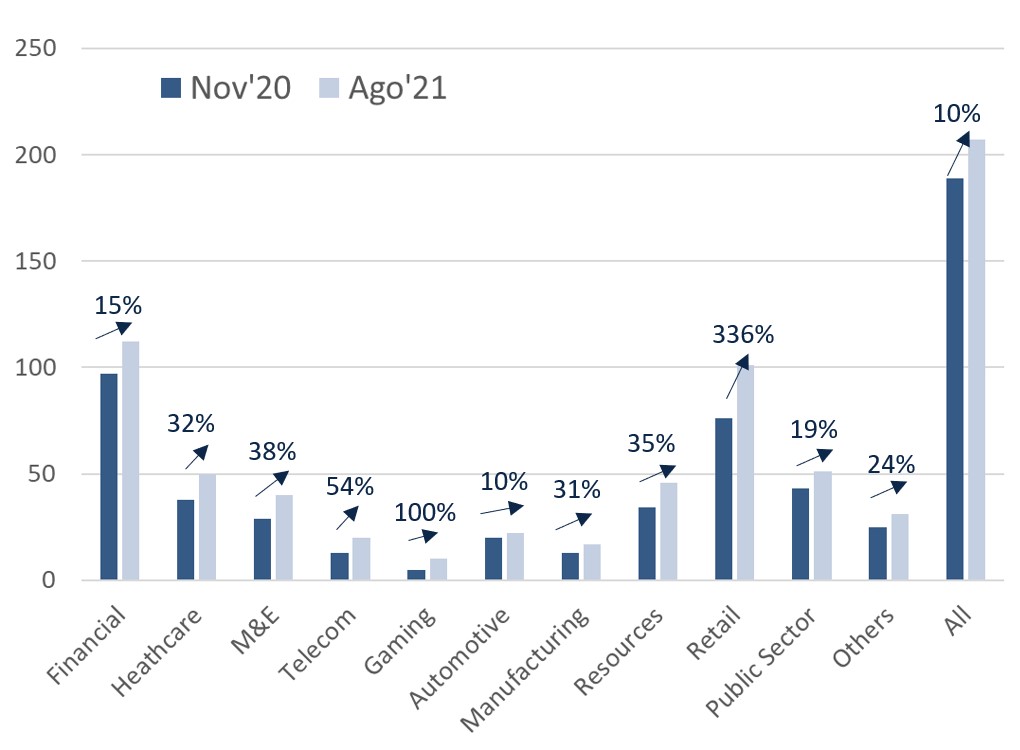}
        %\vspace{-5mm}
        \caption{Nº sellers by category}
        \label{SubFigAWSVendorsEvolution}
    \end{subfigure}
    %\vspace{-3mm}
    \caption{Evolution of AWS' offer from Dec'20 to Mar'21}
    \label{Fig:AWSEvolutionInTime}
\end{figure}

We also conducted a temporal analysis of AWS DM.
Figure~\ref{SubFigAWSProductEvolution} shows how the number of data products offered by AWS in each category evolved from November 2020 to August 2021. Figure~\ref{SubFigAWSVendorsEvolution} shows the evolution of the number of sellers in each category for the same period. From these figures \emph{we see a significant growth taking place in AWS DM} in terms of both data products, which increased a 3\% compound monthly growth rate (CMGR) in this period, as well as sellers, with a 1\% CMGR in the same period. With regards to growth by category, \textit{Gaming} (175\%), \textit{Automotive} (112\%) \textit{Public sector} (79\%), \textit{Financial} (55\%), and \textit{Manufacturing} (48\%) exhibit the highest relative growth in terms of number of products. Moreover, \textit{Financial} and \textit{Public sector} accounted for the highest absolute growth in number of products, adding 444 and 349 respectively in this period.

%\vspace{-3mm}
\section{Comparing data products across marketplaces}
\label{sect:CrossDMcomparison}
Comparing information about data products from different DMs is not a straightforward task since they provide metadata of different granularity and level of detail. In addition, they are using different categorisation systems (often hierarchical ones) to describe their products. To overcome these challenges, we developed a methodology to homogenize the collected data in order to be able to compare similar data products across marketplaces.

\subsection{Dealing with different levels of detail} \label{sect:DealingWithDifferentLevelOfDetail}

Some DMs provide more information than others about their offers. To sort this out, we built a common cross-DM database utilizing a superset of all the different description fields found in different entities. Apart from their category and text descriptive fields, data product records include the time scope, the volume and units, any potential limitations (e.g. maximum number of users, etc), add-ons, granularity of the information, geo-scope at country level, data delivery methods, update frequency and data format.

We normalised and stored in this cross-DM database all the information from the scraped datasets. We managed to fully automate the extraction of most of the fields (18 out of 27), which were directly scraped from the web pages of the different DMs. This extraction was semi-automated for 5 fields, meaning that they were automatically extracted for certain DMs, or retrieved from product descriptions for others, in a process that required a manual check afterwards. For example, this was the case with the field concerning the \emph{update rate} of data, which is usually included in the general description of a data product. Although the presence of the word `\textit{monthly}' may point to a monthly update rate, it was referring to something else in some cases. Information about data volume or data subject units was automatically extracted only for DataRade and BookYourData, and required computer-aided manual typing in the rest of the DM (we highlight and extract numbers and context in descriptions).

%\vspace{-3mm}
\subsection{Dealing with different categorisation systems} \label{sect:DealingWithDifferentCategorization}

Every DM has its own way to classify data. For example, AWS tags data products in 10 different categories, whereas DataRade allows data products to be positioned in a hierarchy of more than 300 categories and more than one (out of 150) use cases.

Furthermore, boundaries between tags are often blurry, and the criteria followed by different DMs to label a data product with a certain category tag are not necessarily coherent. For example, only certain DMs mark `\textit{credit card transaction}' data products as `\textit{financial}', whereas all DMs label them as related to `\textit{marketing}'. Thus, even if we find apparently comparable categories between or across different DMs, we may miss relevant data products due to inconsistencies in the categorization processes that different DMs are utilising.

We addressed the above issues by developing a series of natural language processing (NLP) naïve bayes (NB) classifiers~\cite{Domingos97, Denoyer04, Krishaveni16}. In our first attempt, we wanted to identify similar data products -- data products that belong to the same category -- between two different (source and destination) DMs. As a result, we trained both multinomial and complement versions of NB classifiers to detect data products from the source DM that belong in a certain category by using feature vectors based on the information provided by the data product description from the source DM. We used bag of words~\cite{Ko12} and data preprocessing steps such as removing stop words, and words with numbers, stemming, and TF-IDF transformation~\cite{Salton88, Matic20}. Then we validated the resulting classifier against a manually labelled sample of products from the destination DM.

We utilised the above methodology to build different classifiers to help us compare data products between the two DMs including more price references, namely DataRade (destination DM) and AWS (source DM). We generated our feature vectors based on AWS data product descriptions (source DM) and applied the resulting classifiers to DataRade data products (destination DM). We focused on the two most popular categories: `\textit{Retail, Location and Marketing}' and `\textit{Financial}'. What we are interested to find out is: (1) what percentage of products from those categories can we identify in DataRade, (2) whether the most expensive data products in both DMs , and (3) whether we can enrich our metadata features by adding AWS's inferred categories to products from other DMs. 

We utilised our cross-DM database to generate the train/test datasets at 80/20 split in order to train and test the corresponding classifiers. We observed that multinomial classifiers outperformed the complement NB for this task so we proceeded with the former ones. The resulting classifiers yield an acceptable $F_1$ score above 0.85 (average for 50 executions with different random 80/20 train/test splits). In fact, they identified meaningful and reasonable stems when tagging products related to each category, such as: \\
\noindent\textbf{Financial: }`system', `sec', `exchang', `type', `file', `form', `edgar', `secur', `act’, and `compani'.\\
\noindent\textbf{Retail, Location and Marketing: }`locat', `topic', `b2b', `score', `echo', `trial', `compani', `visit', `intent', `consum'.

However, such classifiers did not perform so well on the validation set, and achieved only $F_1$ scores of 0.73 and 0.43 for `\textit{Financial}' and `\textit{Retail, Location and Marketing}' data. To generalize further our methodology and improve its accuracy, we enriched the train/test datasets with information from more marketplaces. In particular:

\noindent (1) The \textbf{\textit{Financial}} classifier was trained with 95,208 labelled descriptions of data products from 4 different entities (Advaneo, Carto, AWS, and Refinitiv), including 45,298 financial products. 

\noindent (2) The \textbf{\textit{Retail, Location and Marketing}} classifier was trained with 3,828 descriptions from 4 entities (AWS, BookYourData, USASalesLeads, and TelephoneLists), including 1,614 `\textit{Retail, Marketing and Location}' related products.

\noindent (3) The validation set includes 745 manually pre-labelled with both `\textit{Financial}' and `\textit{Retail, Location and Marketing}' tags.

\begin{table}[t]
\caption{Score of NB-Multinomial models} 
%\vspace{-4mm}
\label{tab:ClassifierModelsAccuracy}
\begin{center}
\begin{tiny}
\resizebox{\columnwidth}{!}{%
\begin{tabular}{l|c|c|c|c}
\toprule
\hline
&\textbf{Accuracy}&\textbf{Precision}&\textbf{Recall}&\textbf{$F_1$ Score}\\
%\midrule
\hline
\textbf{Test - Financial}&0.93&0.97&0.81&0.88\\ \hline
\textbf{Test - Retail}&0.95&0.96&0.88&0.91 \\ \hline
\textbf{Val. - Financial}&0.88&0.77&0.69&0.73\\ \hline
\textbf{Val. - Retail}&0.65&0.84&0.29&0.44\\ \hline
\textbf{Val. - Financial (opt)}&0.89&0.72&0.88&0.79\\ \hline
\textbf{Val. - Retail (opt)}&0.78&0.81&0.68&0.74\\
\hline
\bottomrule
\end{tabular}
}
\end{tiny}
\end{center}
\end{table}

By adding products belonging to the same category from other DMs we observed better balance between precision and recall and overall improvement of model generalisation. We also observed an increase of the $F_1$ score in the test set. Particularly, adding information from Refinitiv improves the $F_1$ score from 0.73 to 0.79. In the case of `\textit{Retail, Location and Marketing}', adding information from specialized marketing DMs (BookYourData and USASalesLeads), drastically improves the $F_1$ score from 0.43 to 0.74. We tested multiple classifiers, with and without stemming, and we found that using word-based instead of stem-based features led in general to more accurate results in both cases (+5\% $F_1$ score). Table~\ref{tab:ClassifierModelsAccuracy} shows the accuracy obtained both for 50 random 80/20 train/test splits of the AWS training set (in the first two rows), and the results for the validation set before the optimization in the third and fourth rows, and after the optimization in the last ones.

%\vspace{-3mm}
\subsection{Price comparison results}

We used the above classifiers to label data products in DataRade in order to obtain the price distribution of these two categories. We located 619 and 701 `\textit{Financial}' and `\textit{Retail, Location and Marketing}' data products in DataRade. They represent 39\% and 44\% of the total sample, respectively. As happened in AWS, not only do those categories contain the largest number of products in this DM, but the most expensive ones are tagged as `\textit{Retail, Location and Marketing}', as well. We built similar classifiers for the rest of AWS data categories and enriched our sample by homogeneously labelling the rest of our data products.

\begin{figure}[t]
    \centering
    \includegraphics[width=0.35\textwidth]{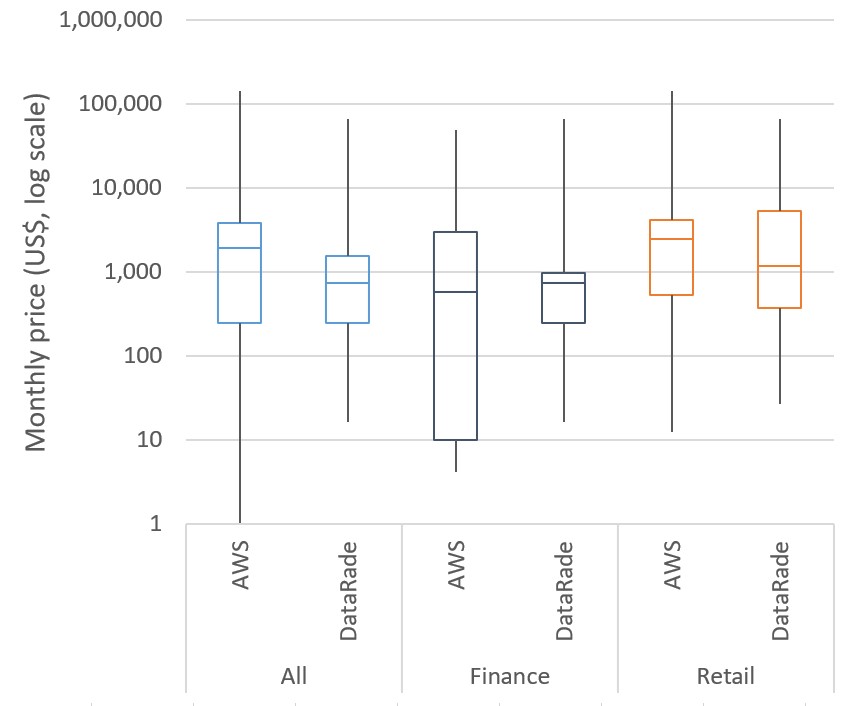}
    %\vspace{-3mm}
    \caption{Box plot of subscription-based data prices in AWS vs DataRade}
    \label{FigCrossCDFAWSDataRade}
\end{figure}

Figure~\ref{FigCrossCDFAWSDataRade} shows a box plot that compares the distribution of prices for `\textit{Financial}' and `\textit{Retail, Location and Marketing}' subscription-based data products in AWS and DataRade. Median prices are significantly higher in AWS (close to $\times 4$) than in DataRade. Products in DataRade start at 16.5 USD/month, while there exist 301 cheaper products in AWS (12,6\% of them). In both DMs, the most expensive data products are related to \textit{Retail, Location and Marketing}, a category whose median price is around 30-60\% higher than the global median. On the contrary, the median prices of financial data products are lower in both DMs (US\$583 and US\$750 per month for AWS and DataRade respectively) are below the median prices for all products (US\$1,927 and US\$750 monthly).

Does this methodology work if we switch source and destination DMs? In order to answer this question, we trained NB classifiers to detect products in AWS related to relevant use cases and categories in DataRade. In this case, DataRade acted as the source DM, i.e., it provided descriptions and tagging information to train the classifiers, whereas AWS' role was the destination DM, whose products we labelled with some of DataRade's tags and driven by the criteria we learnt from the source DM. In particular, we focused on products belonging to the `\textit{B2B Marketing}', `\textit{Audience Targeting}' and `\textit{Risk Management}' use cases in DataRade, some 46, 48 and 30 products out of 745 respectively. Since the training set is imbalanced and the number of samples is low, complement NB outperformed multinomial NB in this case. We trained the classifiers and obtained the log-probability of belonging in each category for all the data products in AWS. As a result, at least 16 out of the top 20 data products showing the highest log-probability turned out to be useful for those specific use cases.

\begin{figure*}[t]
    \centering
    \begin{subfigure}[b]{0.32\textwidth}
        \centering
        \includegraphics[width=\textwidth]{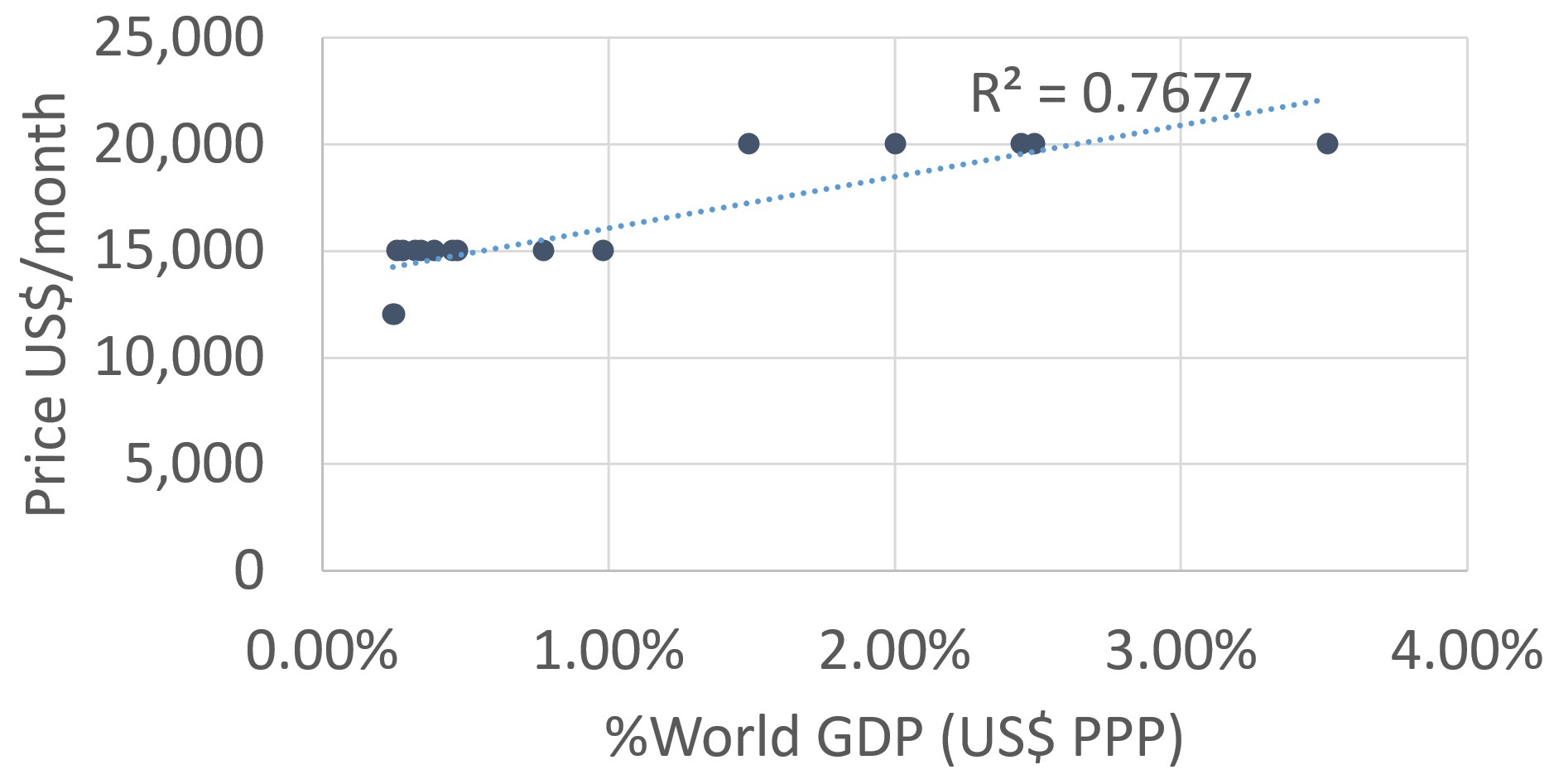}
        %\vspace{-6mm}
        \caption{Mobile coverage}
        \label{SubFigLTECoverage}
    \end{subfigure}
    \begin{subfigure}[b]{0.32\textwidth}
        \centering
        \includegraphics[width=\textwidth]{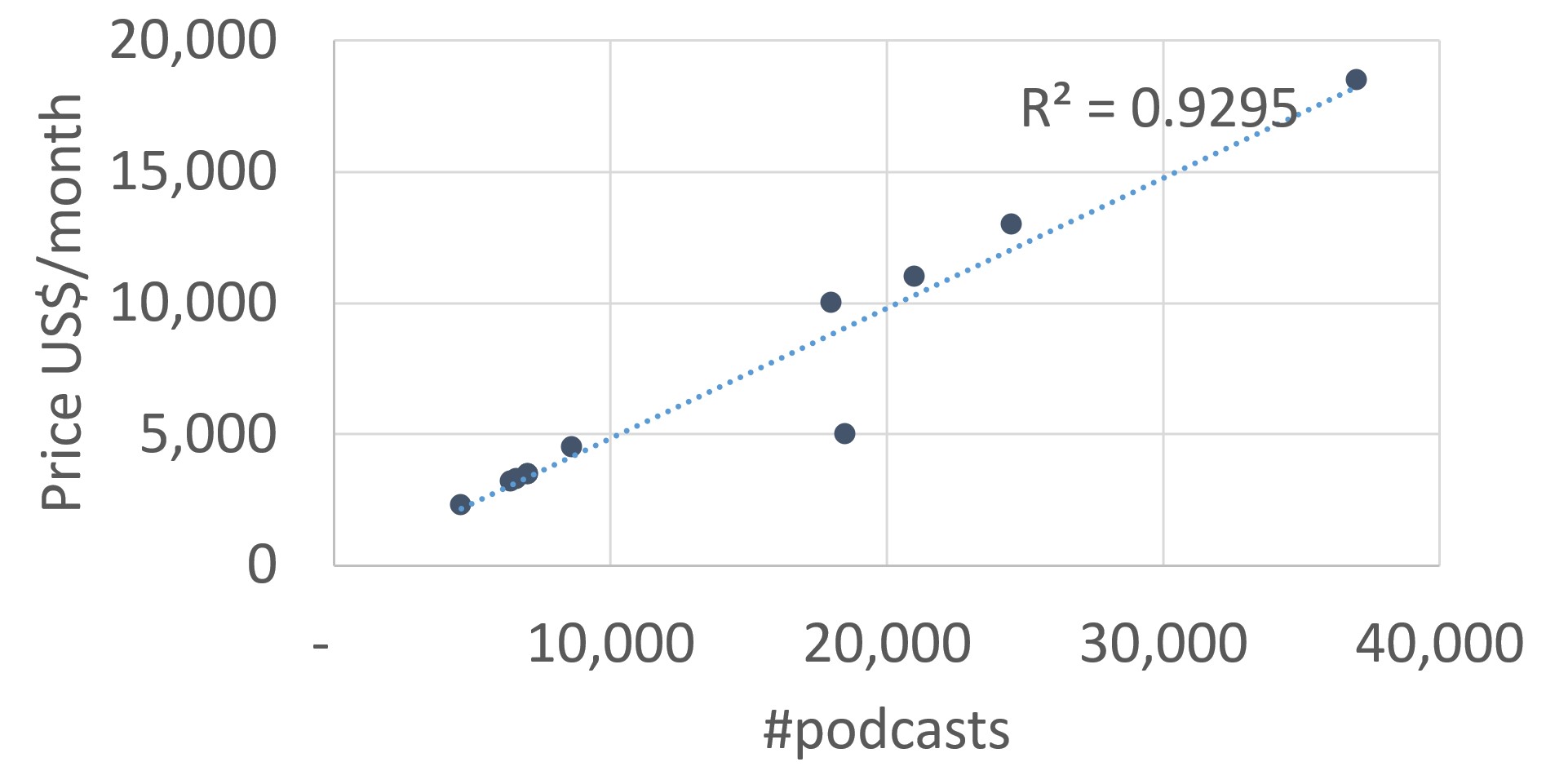}
        %\vspace{-6mm}
        \caption{Podcast metadata information}
        \label{SubFigPodcastMetadata}
    \end{subfigure}
    \begin{subfigure}[b]{0.32\textwidth}
        \centering
        \includegraphics[width=\textwidth]{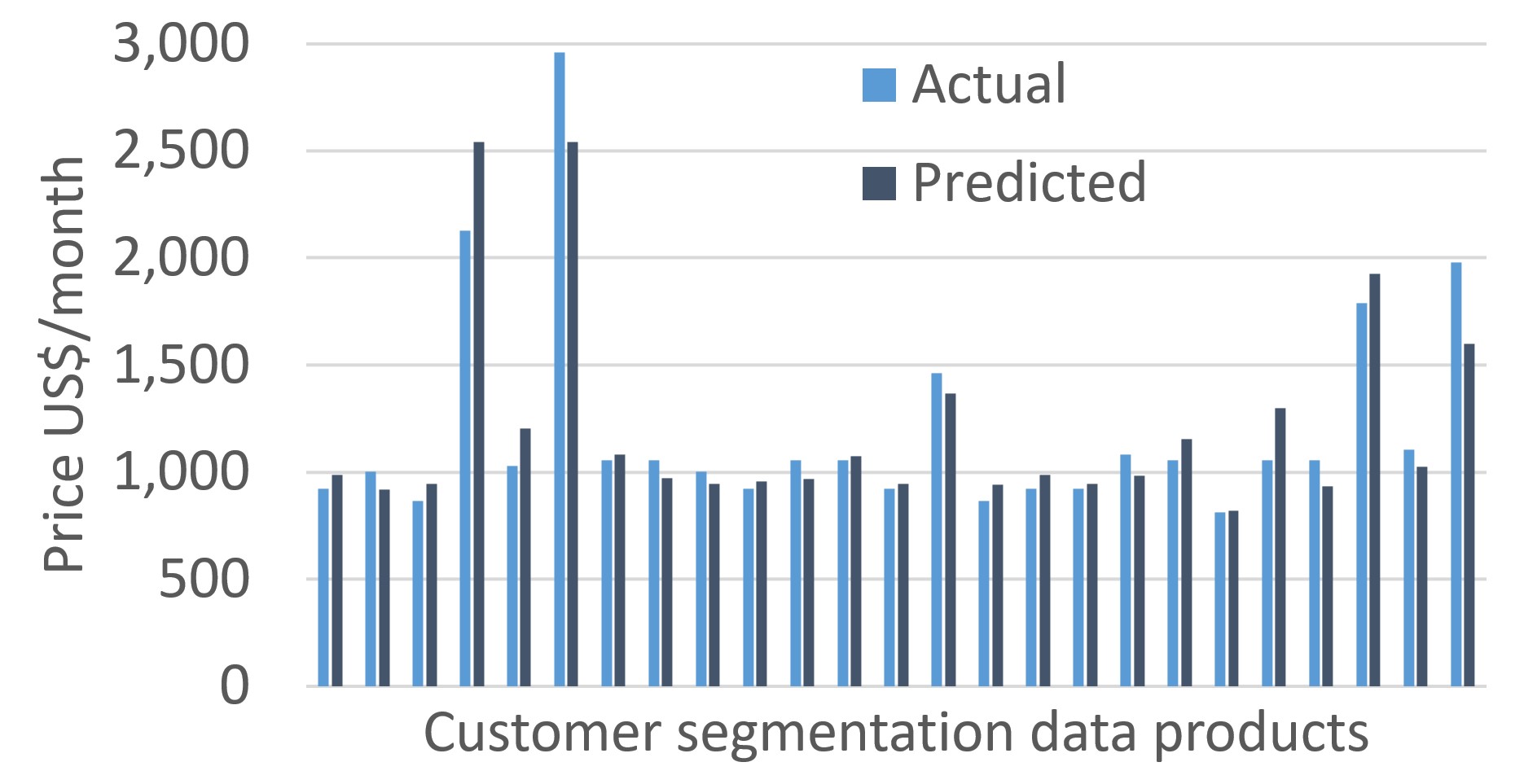}
        %\vspace{-6mm}
        \caption{Customer segmentation products}
        \label{SubFigCustomerSegmProducts}
    \end{subfigure}
    %\vspace{-3mm}
    \caption{Pricing regression examples from specific sellers}
    \label{FigSpecificPricingExamples}
\end{figure*}

%\vspace{-4mm}
\section{Which are the features driving data product prices?} 
\label{sect:UnderstandingFeaturesDrivingDataPrices}
So far we have seen an overview of data product prices, looked at the prices of particular categories, and developed and applied a methodology to homogeneously label data products across DMs in our study. Our final goal is to understand whether there exist any key features that have a determinant role on the price of data products. 

To answer this question, we first inspect manually our cross-DM database to identify any common distinctive features of top most valuable data products. Then we shift our attention to particular data sellers and see how they distinguish their most expensive from their less expensive data products. With these insights, we construct a features dataset to train regression models for predicting the prices of real commercial data products. Finally, we conduct feature importance analysis to answer which features have the highest impact on the observed prices for financial, marketing and healthcare-related data.

%\vspace{-4mm}
\subsection{Understanding the key features of highly priced data products}
\label{sect:KeyFeaturesAffectingTopDatasets}
Section~\ref{sect:Analyzing productsInAsingleDM} presented a first empirical analysis of features of highly priced data products. Having enriched our dataset for constructing the classifier of the previous section, we will now use this extra data to better understand why the top 33 data products are worth more than US\$30,000 per month. Looking at the different collected fields in our database we observe the following:
%\vspace{-2mm}
\begin{itemize}
    \item They all include huge amounts of data from millions of people, tens of thousands of locations or companies, etc.
    \item 20 (61\%) of them offer daily updates.
    \item They provide fresh rather than historical data. Only 4 (12\%) offer over 2 years historical data, and 11 (33\%) of them do not provide past data.
    \item 22 (67\%) of them are US-focused, and 7 (21\%) are global.
    \item 25 (73\%) of them relate to \textit{Retail, Location and Marketing}.
    \item B2B most valuable data products include precise enterprise data, as well as contact information for key people in them.
    \item At least 16 (48\%) of them enable a very granular location-based analysis, and 9 (27\%) of them provide geo-located data.
    \item 7 B2C marketing products (21\%) allow for session reconstruction (i.e., connecting the different data points of individuals/entities).
\end{itemize}

%Even though we identified features that may potentially be related to price, the former analysis is limited to top valuable data products. Next, we will deal with offers from specific sellers in order to identify pricing strategies and the features they rely on.

\begin{table*}[t]
\caption{List of feature groups}
%\vspace{-4mm}
\label{tab:FeaturesAndGroups}
\begin{center}
\begin{tiny}
\resizebox{\textwidth}{!}{%
\begin{tabular}{ll|p{6cm}|c|l}
\toprule
\hline
\textbf{Question}&\textbf{Group}&\textbf{Definition}&\textbf{Nº features}&\textbf{Example of features}\\
\midrule
\hline
\multirow{3}{*}{\textbf{What?}}&\textbf{Category}&Labels attached to the product that define the type of data it contains&11&`Weather', `Gaming', `Financial'\\ \cline{2-5}
&\textbf{Description}&Stem-like features obtained from data product descriptions&up to 2000&`wordmarket', `wordidentifi', `wordlist'\\ \cline{2-5}
&\textbf{Identifiability}&Tells whether the product allows the buyer to recognize the activity of individuals or to identify specific companies&2&`IdSessions', `IdCompanies'\\ \hline
\multirow{2}{*}{\textbf{How much?}}&\textbf{Volume}&Normalized nº units covered broken down by the nature of such units&14&`units', `people', `entities'\\ \cline{2-5}
&\textbf{Update rate}&Defines the frequency between data updates as announced by the seller&11&`realtime', `monthly', `hourly'\\ \hline
\multirow{3}{*}{\textbf{How?}}&\textbf{Delivery method}&Defines how the buyer can have access to data&8&`S3Bucket', `Download', `FeedAPI'\\ \cline{2-5}
&\textbf{Format}&Defines the way in which data is arranged&17&`txt', `shapefile', `xls'\\ \cline{2-5}
&\textbf{Add-ons}&Tells whether the product attaches any add-on or has any limitations&2&`ProfServices', `Limitations'\\ \hline
\textbf{When?}&\textbf{History}&Time scope included &1&`History'\\ \hline
\textbf{Where?}&\textbf{Geo scope}&Metrics about countries included in the data product&up to 249&`NCountries', `USA', `Canada'\\
\hline
\bottomrule
\end{tabular}
}
\end{tiny}
\end{center}
\end{table*}

%\vspace{-3mm}
\subsection{Seller-specific pricing strategies}
\label{sect:specificPricingExamples}

The above observations provide valuable insights for constructing features that may correlate strongly with price. Since they capture only the top-most highly-priced data, we also looked from the perspective of sellers, and collected some of their pricing strategies. Looking at \emph{specific} sellers we found that surprisingly simple regression models relying on specific features of each seller's data products were able to accurately predict their prices. Figure~\ref{FigSpecificPricingExamples} depicts three such examples for telecom, recommender systems, and customer segmentation data. From these examples, we observe that: 
%\vspace{-1mm}
\begin{itemize}
    \item A seller offering mobile network infrastructure and coverage data by country, by grouping products in a few price tiers that depend on their gross domestic product (GDP) (see Fig.~\ref{SubFigLTECoverage}).
    \item A seller offering metadata about podcasts on the Internet uses language to segment their products, whose price is almost proportional to the number of podcasts they include (US\$0.5 per podcast, see Fig.~\ref{SubFigPodcastMetadata}).
    \item A well-known leader in the consumer segmentation data market segment relies on the population covered, its purchase power, and the granularity of the information provided to set the prices of its products by country. Figure~\ref{SubFigCustomerSegmProducts} compares the prices predicted by our model to the actual prices and shows a more than acceptable accuracy ($R^2$ score = 0.88, mean absolute error (MAE) = 10\% of the average, mean relative error (MRE) = 9\%).
\end{itemize}
%\vspace{-1mm}
%In fact, we achieved a $R^2 = 0.88$ score when predicting the price of their products with a simple linear regression model fed with 25 consumer segmentation products available for different countries. The model just considered the population covered, the percentage of GDP of the country, the level of granularity meaning the number of geographic areas for which average information is reported, and the average population of such areas as independent variables. Subfigure .

So far we have spotted specific features that seem to be relevant for top-valued data products, and discovered other features that support pricing strategies from specific sellers. Are these observations supported by our data as a whole? Are we able to train regression models that are not only applicable to specific sellers but to the whole base of data products? Before that, we need to tweak our database in order to quantify such features and build a training set for such regressors. In the next section, we will provide some insights about this process and evaluate the linear correlation of individual features with respect to prices.

%\vspace{-3mm}
\subsection{A feature matrix for regression models}
\label{sect:PriceCorrelationIndividualFeatures}
An additional preprocessing step is needed in order to transform the fields of our cross-DM database into a set of valuable features that can be ingested by ML regression algorithms. This process uses the NLTK~\cite{Bird09} and Scikit-learn~\cite{Pedregosa11} Python libraries and includes mainly the following steps:

\noindent(1) Extraction of `\textit{word}' features from the title and the textual description of each data product. We use bag of words~\cite{Ko12} and data preprocessing steps such as removing stop words and words with numbers, TF-IDF transformation~\cite{Salton88}, and stemming~\cite{Matic20}. In addition, we have sellers' names removed from the vocabulary, so as to avoid bias introduced by knowing their identity. Finally, we prepare matrices for different vocabulary lengths to be able to optimize each algorithm for this parameter.

\noindent(2) Breakdown of volume-related fields in 13 different groups depending on their nature. For example, we separate data products targeting `entities or `companies, from those whose subjects are `individuals in different features. The resulting comparable units are in turn normalized, and a new overarching feature (`units') measuring the percentage of units covered is added to compare data products across groups of units. 

\noindent(3) Calculation of country-level binary features to indicate whether a certain country is covered by a data product.

\noindent(4) Homogenization of the units of time when measuring the time scope of the products, what we will call \textit{history}. 

As a result of this \textit{featurization} process, we reduce each sample product to a feature vector and produce a feature matrix to train our regression models. 

Moreover, we organize features in 10 disjoint sets according to their nature and the basic questions they answer about data products. These groups contain features related to the description of data products, their units or volume, their time, geographical scope, update rate, categories, delivery methods, format, whether they allow to reconstruct sessions of data subjects, and whether add-ons are included in the product. Table~\ref{tab:FeaturesAndGroups} shows the list of feature groups and some examples of their individual features.

Our challenge now is measuring which features and groups of features are more significant in determining the price of data products in commercial marketplaces. Moreover, we carry out such analysis three times in order to compare the results for financial, marketing and healthcare-related data, and once more including all the products in our sample. Before feeding the models, we reduce the number of input features by discarding features that have a unique value, which may appear when filtering the complete dataset by \textit{category}. Next, we unify groups of features showing a high cross-correlation among them, i.e., $R^2 \geq 0.9$. We also evaluate the linear correlation of individual features with respect to data product prices. Not surprisingly, it turns out that none of them is linearly correlated to price, and their $R^2$ score is always below 0, as opposed to what we found for specific sellers. 

\begin{table*}[t]
\caption{Top 10 most relevant features by category and algorithm}
%\vspace{-3mm}
\begin{tiny}
\label{tab:Top10MostRelevantIndividualFeatures}
\resizebox{\textwidth}{!}{%
\begin{tabular}{c|c|c||c|c|c||c|c|c||c|c|c}
\toprule
\hline
\multicolumn{3}{c||}{\textbf{Financial}}                          & \multicolumn{3}{c||}{\textbf{Marketing}}                          & \multicolumn{3}{c||}{\textbf{Healthcare}}
    & \multicolumn{3}{c}{\textbf{All}} \\ \hline
\textbf{RF} & \textbf{kNN} & \multicolumn{1}{c||}{\textbf{GBR}} & \textbf{RF} & \textbf{kNN} & \multicolumn{1}{c||}{\textbf{GBR}} & \multicolumn{1}{c|}{\textbf{RF}} & \multicolumn{1}{c|}{\textbf{kNN}} & \textbf{GBR} & \multicolumn{1}{c|}{\textbf{RF}} & \multicolumn{1}{c|}{\textbf{kNN}} & \textbf{GBR}\\ \hline
units&units&units&units&units&csv&units&csv&wordlist&units&units&DelMethod\\ \hline
entities&Email&S3Bucket&locationdata&History&units&people&units&DelMethod&yearly&IdSessions&S3Bucket\\ \hline
S3Bucket&Download&wordmonthli&Weight&USA&yearly&wordhealth&daily&wordhospit&Download&Retail&units\\ \hline
wordsubmit&daily&wordstock&USA&IdSessions&people&wordtrend&wordmarket&wordidentifi&wordreport&USA&entities\\ \hline
Download&IdCompanies&worddeliv&IdCompanies&NCountries&RESTAPI&wordmedic&wordgo&wordamerica&entities&IdCompanies&requests\\ \hline
people&USA&people&txt&Financial&wordqualiti&wordglobal&Limitations&wordhealth&people&worduser&people\\ \hline
txt&wordmarket&DelMethod&daily&Others&wordaccur&csv&locationdata&wordreport&wordmarket&Others&yearly\\ \hline
wordedgar&Retail&txt&S3Bucket&people&wordidentifi&DelMethod&wordpopul&wordstudi&monthly&wordconsum&pdf\\ \hline
wordcustom&wordcontact&wordneed&wordmonthli&wordcontact&wordwebsit&wordinsight&wordprofil&wordupdat&wordcontact&Canada&RESTAPI\\ \hline
wordlist&realtime&wordsubmit&wordvp&Email&UIExport&wordreport&wordinsight&wordcontact&wordwebsit&wordcompani&wordlist\\ \hline

\bottomrule
\end{tabular}
}\end{tiny}
\end{table*}

%\vspace{-3mm}
\subsection{Analyzing feature importance for regression models}
\label{subsect:FeatureImportanceForRegressionModels}
%In the last section we tried unsuccessfully to detect meaningful linear correlations of individual features with respect to price as we did with products of specific sellers. 
Regression models can be used for feature importance analysis. In this section we use a range of such techniques to understand which ones have the higher impact on data product prices. 

Owing to their stochastic nature, training several regression algorithms and comparing their outcomes is key to obtaining robust conclusions. Consequently, we tested 9 different regression models, and selected 3 of them, namely Random Forest~\cite{Breiman01} (hereinafter, RF), k-Nearest Neighbours~\cite{Kramer11} (hereinafter, kNN), and Gradient Boosting regressors~\cite{Friedman00, Mason99} (hereinafter, GBR) to be further optimized. We carried out this process separately for financial, marketing, healthcare-related and all data products in our sample, as detailed in the appendix~\ref{appendix:RegressionModels}. 

As a result, we obtained at least one model that achieves a $R^2$ score of 0.78 by category and accurately fit the prices of data products ($MAE \leq 0.25$ and $MSE \leq 0.16$). We ran 4 different feature importance analysis on top of those optimized models, as explained in the appendix~\ref{appendix:FeatureImportanceAnalysis}.

%\vspace{-2mm}
\subsubsection{Analyzing the importance of individual features}
\label{sect:AnalyzingIndividualFeatureImportance}
We have found that 50\% of the positive LOO (see~\ref{appendix:RegressionModels} for details) and 67\% of the $\Delta R^2$ score by permuting values owe to the top 10 most relevant features on average for specific categories of data. Table~\ref{tab:Top10MostRelevantIndividualFeatures} lists these features in descending order of importance. Note that we would need more than 25 features to achieve equivalent scores if we include all the products. Next we provide some details about the most important features of each specific category of data:

\textbf{Financial:}
`\textit{Units}' and `\textit{entities}' are by far the most relevant features when determining prices of financial products. Not only do volume-related features rank number one, but they are on average $4$ times more important than the second feature for RF and GBR. Other features relate to specific characteristics of financial data products according to their category (e.g., `\textit{Retail}') or their description. For instance, RF relies on the stem `\textit{edgar}', which stands for SEC’s Electronic Data Gathering, Analysis, and Retrieval System. The stem `\textit{custom}' refers to the valuable possibility of personalizing data products (e.g., select which companies we want financial data from). Features related to delivery methods (e.g., `\textit{S3bucket}' or `\textit{Download}') and update rate (e.g., `\textit{realtime}' or `\textit{daily}') stand out in terms of relevance, as well.

\textbf{Marketing:} With regards to marketing data products, features related to units, such as `\textit{units}' and `\textit{people}' rank high, as well. Specific characteristics of data play a relevant role, too. For example, stems like `\textit{contact}' are used to locate contact lists, a family of B2C and B2B marketing products, the stems `\textit{qualiti}' and \textit{accur} refer to the high-quality and accuracy of data, as advertised by sellers. Interestingly, location data seems to be important in this case, as the presence of `\textit{locationdata}' feature suggests. So does scope: `\textit{USA}' and `\textit{NCountries}' (number of countries) seem to play a key role in RF and kNN results. Finally, the features `\textit{IdSessions}' and `\textit{IdCompanies}' indicates that being able to reconstruct sessions of anonymized individuals and being able to identify merchants are price drivers for marketing products. 

\textbf{Healthcare:} The `\textit{what}' seems to be more important than the `\textit{how much}' when pricing healthcare products, as suggested by the amount of stem-related features in the top-10. This is due to the heterogeneity of data products belonging in this category, ranging from contact lists of healthcare practitioners, and companies to training datasets for medical / pharmaceutical ML translators, or reports on specific medications. Stems like `\textit{trial}' or `\textit{studies}' help in identifying what a dataset refers to. For instance, the stem `\textit{go}' refers to an official check-in and rating system in the US to limit the spread of COVID. In this case, volume-related features seem to be important for two algorithms only. Features related to the update rate, data format (`\textit{csv}'), the number of available delivery options (`\textit{DelMethod}') and the presence of `\textit{Limitations}' (e.g., limited number of reports or data exports included) seem to be also relevant to determine product prices, too.

%\vspace{-2mm}
\subsubsection{Analyzing the importance of groups of features} 
\label{sect:AnalyzingGroupFeatureImportance}
Since LOO is often negligible for individual features, we have repeated this analysis by removing groups of features, and calculated the Shapley value as explained in appendix~\ref{appendix:FeatureImportanceAnalysis}. Whereas LOO measures gains or loses in accuracy of a model when features belonging in a group are removed from the input matrix, Shapley values better capture the complementarity among groups and take into consideration their individual predictive power, as well. Table~\ref{tab:RankingFeatureGroupsLOO} and Table~\ref{tab:ShapleyValueFeatureGroups} list the LOO and the Shapley values by group of features in descending order of importance.

\begin{table}[t]
\caption{LOO values by feature group}
\label{tab:RankingFeatureGroupsLOO}
\resizebox{\columnwidth}{!}{%
\begin{tabular}{l||c|c|c||c|c|c||c|c|c||c|c|c}
\toprule
\hline
\multicolumn{1}{c||}{\multirow{2}{*}{\textbf{Group}}} &
  \multicolumn{3}{c||}{\textbf{Financial}} &
  \multicolumn{3}{c||}{\textbf{Marketing}} &
  \multicolumn{3}{c||}{\textbf{Healthcare}} &
  \multicolumn{3}{c}{\textbf{All}} \\ \cline{2-13} 
\multicolumn{1}{c||}{} &
  \textbf{RF} &
  \textbf{kNN} &
  \multicolumn{1}{c||}{\textbf{GBR}} &
  \textbf{RF} &
  \textbf{kNN} &
  \multicolumn{1}{c||}{\textbf{GBR}} &
  \multicolumn{1}{c|}{\textbf{RF}} &
  \multicolumn{1}{c|}{\textbf{kNN}} &
  \textbf{GBR} &
  \multicolumn{1}{c|}{\textbf{RF}} &
  \multicolumn{1}{c|}{\textbf{kNN}} &
  \textbf{GBR} \\ \hline \hline
\textbf{Descriptions}&0.027&0.025&0.066&0.021&0.034&0.098&0.054&0.425&0.052&0.023&-0.020&0.079\\ \hline
\textbf{Volume}&0.092&0.182&0.167&0.171&0.138&0.199&0.048&0.014&0.052&0.138&0.123&0.142\\ \hline
\textbf{Scope}&-0.005&-0.007&-0.001&-0.003&-0.006&0.000&0.015&0.000&-0.011&-0.003&-0.002&0.000\\ \hline
\textbf{Del. method}&0.005&0.032&0.011&0.000&0.018&0.008&0.019&0.017&0.003&0.002&0.010&0.008\\ \hline
\textbf{Format}&0.002&0.004&0.010&0.007&0.001&0.023&0.007&0.030&0.000&0.002&0.007&0.006\\ \hline
\textbf{Category}&-0.002&0.001&0.001&-0.001&-0.003&0.001&0.013&-0.033&-0.006&0.001&0.000&0.003\\ \hline
\textbf{Add-ons}&-0.001&0.007&-0.001&-0.001&0.000&0.001&0.000&0.022&0.000&0.001&0.001&0.000\\ \hline
\textbf{Identifiability}&-0.002&0.016&0.002&-0.001&0.006&0.004&0.010&0.000&-0.009&0.000&0.008&0.000\\ \hline
\textbf{History}&-0.001&0.000&0.000&-0.003&0.004&0.000&0.009&0.000&0.000&0.002&0.000&-0.001\\ \hline
\textbf{Update Rate}&0.001&0.023&0.001&0.036&0.000&0.016&0.010&0.021&0.000&0.021&-0.002&0.014\\ \hline

\bottomrule
\end{tabular}
}
\end{table}

\begin{table}[t]
\caption{Shapley values by feature group} 
%\vspace{-4mm}
\label{tab:ShapleyValueFeatureGroups}
\resizebox{\columnwidth}{!}{%
\begin{tabular}{l||c|c|c||c|c|c||c|c|c||c|c|c}
\toprule
\hline
%&\multicolumn{3}{c}{Financial}&\multicolumn{3}{c}{Marketing}&\multicolumn{3}{c}{Healthcare}\\
%&RF&kN&GB&RF&kN&GB&RF&kN&GB\\
\multicolumn{1}{c||}{\multirow{2}{*}{\textbf{Group}}} &
  \multicolumn{3}{c||}{\textbf{Financial}} &
  \multicolumn{3}{c||}{\textbf{Marketing}} &
  \multicolumn{3}{c||}{\textbf{Healthcare}} &
  \multicolumn{3}{c}{\textbf{All}} \\ \cline{2-13} 
\multicolumn{1}{c||}{} &
  \textbf{RF} &
  \textbf{kNN} &
  \multicolumn{1}{c||}{\textbf{GBR}} &
  \textbf{RF} &
  \textbf{kNN} &
  \multicolumn{1}{c||}{\textbf{GBR}} &
  \multicolumn{1}{c|}{\textbf{RF}} &
  \multicolumn{1}{c|}{\textbf{kNN}} &
  \multicolumn{1}{c||}{\textbf{GBR}} &
  \multicolumn{1}{c|}{\textbf{RF}} &
  \multicolumn{1}{c|}{\textbf{kNN}} &
  \textbf{GBR} \\ \hline
%\midrule
\hline
\textbf{Description}&0.155&0.266&0.222&0.247&0.153&0.152&0.232&0.290&0.236&0.113&0.176&0.187\\ \hline
\textbf{Volume}&0.211&0.216&0.184&0.290&0.241&0.241&0.168&0.125&0.131&0.211&0.210&0.174\\ \hline
\textbf{Format}&0.087&0.006&0.086&0.027&0.046&0.094&0.090&0.077&0.082&0.072&0.087&0.071\\ \hline
\textbf{History}&0.072&0.000&0.059&0.009&0.037&0.036&0.063&0.001&0.046&0.058&0.010&0.037\\ \hline
\textbf{Update Rate}&0.088&0.056&0.084&0.060&0.032&0.050&0.046&0.145&0.041&0.067&0.034&0.067\\ \hline
\textbf{Del. Method}&0.036&0.054&0.044&0.093&0.075&0.049&0.030&0.040&0.035&0.062&0.062&0.074\\ \hline
\textbf{Identifiability}&0.034&0.038&0.028&0.052&0.027&0.048&0.040&0.001&0.031&0.056&0.022&0.039\\ \hline
\textbf{Scope}&0.056&0.046&0.050&0.032&0.044&0.036&0.030&0.001&0.040&0.061&0.015&0.024\\ \hline
\textbf{Type}&0.071&0.021&0.044&0.018&0.043&0.037&0.017&0.031&0.039&0.070&0.063&0.055\\ \hline
\textbf{AddOns}&0.021&0.003&0.021&0.012&0.028&0.038&0.048&0.053&0.041&0.055&0.026&0.045\\ \hline

\hline
\bottomrule
\end{tabular}
}
\end{table}

\begin{figure}[t]
    \centering
    %\vspace{-3mm}
    \includegraphics[width=0.48\textwidth]{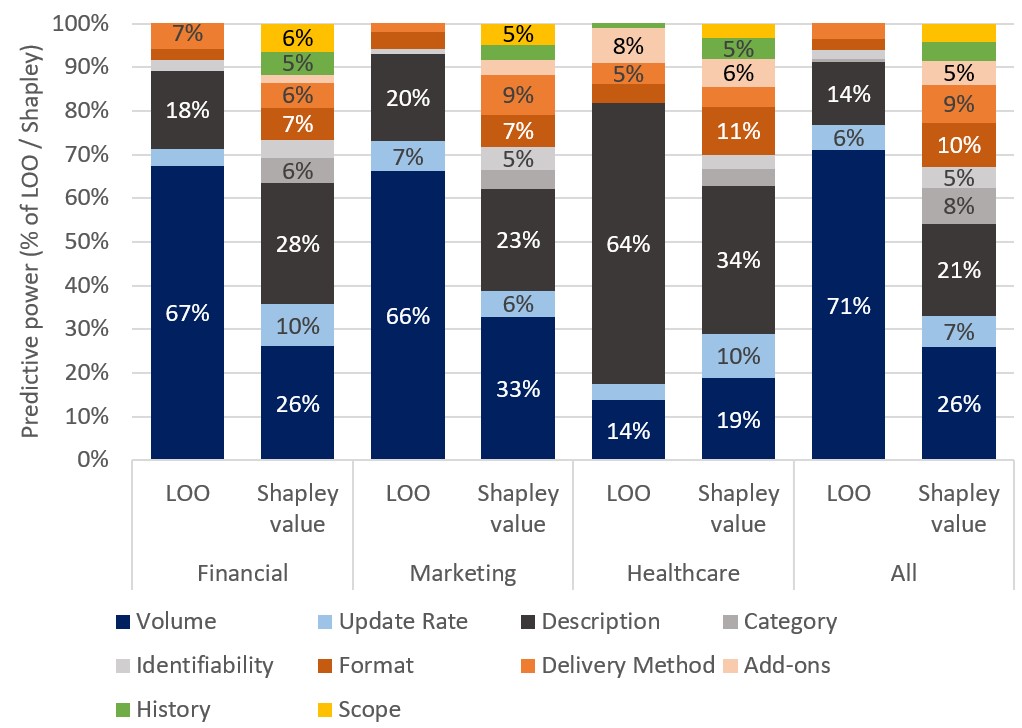}
    %\vspace{-3mm}
    \caption{Predicting power of feature groups by category}
    \label{FigPredictingPowerByCategoryAndGroup}
\end{figure}

Figure  ~\ref{FigPredictingPowerByCategoryAndGroup} plots the percentage of the sum of Shapley and LOO values that each group of features represents, what we call their \textit{predictive power}, and illustrates how important each group is for determining the prices of each category of data products. We have piled together and colored in gradients groups responding to the same question.

Note that the algorithms, in the absence of certain features, will try to replace or infer them through other features in order to come up with the best estimation possible. We have observed that this happens with `\textbf{category}' labels or `\textbf{add-ons}' and product descriptions, and it is also the reason why LOO values are generally smaller than Shapley values.

By looking at Fig.~\ref{FigPredictingPowerByCategoryAndGroup}, we can confirm that features related to `\textbf{volume}' and `\textbf{descriptions}' are the most relevant groups driving data prices: at least half of the predictive power owes to those two groups of features according to their Shapley values. However, their mix differs across categories. While `\textit{volume}' is definitely the most relevant group for marketing data products, it is not so relevant for healthcare-related data due to the heterogeneity and less sensitivity to volume of products belonging in this category.

Data `\textbf{update rate}' and its '\textbf{format}' are consistently relevant across all data categories, but to a lesser extent (6-11\% of the prediction score), whereas the Shapley values of the other groups differ across categories: `\textbf{history}' (meaning the time span of data delivered) is more relevant for financial and healthcare-related data, `\textbf{delivery methods}' are more relevant for marketing data, and ´\textbf{identifiability}' is important in general, but especially for marketing products.

These results are in line with our discussion based on the relevance of individual features in the previous section.

In summary, it is mostly `\textit{what}´, as captured in product description and categories, and `\textit{how much}´ data is being traded that determine the prices of data products. Since relevant descriptive features are diverse and strongly differ across data categories, we failed to find a single feature other than `\textit{units}´ (with exceptions, such as healthcare-related data) that consistently shows a significant \textit{predictive power}. `\textit{How}´ data is delivered to buyers proved to be important too, and accounts for 15-24\% of \textit{predictive power} according to Shapley. Finally, historical time span (`\textit{when}') and geographical scope (`\textit{where}') of data products, whose score oscillates around 5\% for every data category, are apparently less relevant in driving prices.

%\vspace{-3mm}
\section {Related works}
\label {sect:RelatedWorks}
Even though several surveys related to data marketplaces have been recently published~\cite{Schomm13,Stahl14,Spiekermann19}, our work is, to the best of our knowledge, the first empirical measurement study that deals with the prices of data products sold in commercial data marketplaces.

In fact, the lack of empirical data around dataset prices is considered as a key challenge in data pricing research~\cite{Pei20}. According to some authors, some techniques to set the prices of digital products~\cite{Shapiro00} or cloud services~\cite{Wu19} are applicable to data products, as well. Some authors proposed auction designs to set the prices of digital goods and data products~\cite{Goldberg01, Goldberg03}. Novel AI/ML data marketplace architectures have been proposed under the concept of value-based pricing~\cite{Ohrimenko19, Agarwal19, Chen19}. Moreover, some authors defined pricing strategies and DMs based on differential privacy~\cite{Ghosh11, Li15} or queries to a database~\cite{Koutris12, Chawla19}. All of them work on analysing the theoretical properties for fair, arbitrage-free pricing, but leave the responsibility of actually defining absolute prices to both buyers and sellers. Quality-based pricing~\cite{Heckman15} is the one closest to our approach. According to it, the value of data must be assessed by evaluating and assigning weights to certain quality features. Even though some additional works have provided data pricing strategies for sellers based on this idea~\cite{Yu17}, we are not aware of any measurement study that has been able to derive weights for such features from real data.

The pricing of personal data of individuals has received some attention from the privacy and measurement community. There are measurement studies based on prices carried over the Real Time Bidding protocol~\cite{Olejnik14, Papadopoulos17} as well as more traditional survey-based studies~\cite{Carrascal13}. These works report prices for the data and the attention of individuals and, therefore, have nothing to do with B2B datasets traded in modern DMs. 

Finally, cross-marketplace analysis and discoverability of data has been pointed out as a significant challenge by data marketplace vision papers~\cite{Fernandez20}. Google Dataset Search has proposed a standard for providing metadata for their crawlers~\cite{Brickley19}. Discoverability of data is the \textit{leit motiv} of data marketplaces and data aggregators, such as DataRade. Discoverability initiatives do not touch upon pricing questions.

%However, due to the current fragmentation of data markets, there is a need of an overarching solution not only to discover data, but to provide transparency on how much a piece of data might be worth in the market across DMs, and why. 

%\vspace{-3mm}
\section{Conclusions and future work}
\label{sec:conclusions}

Our work has provided a first glimpse into the growing market for B2B data. Despite having worked in a range of pricing topics in the past, prior to conducting this study, we did not have the slightest idea even for fundamental questions such as ``What are typical prices for data products sold online?'', or ``What types of data command higher prices?''. Our work has produced answers to those and many other questions. We have seen that while the median price for data is few thousands, there exist data products that sell for hundreds of thousands of dollars. We have also looked at the categories of data and the specific per-category features that have the highest impact on prices. Having scraped metadata for hundreds of thousands of data products listed by 9 real-world DMs and 27 DPs we found fewer than ten thousand that were non-free and included prices. We believe that this is due to the fact that in many cases prices are left to direct negotiation between buyers and sellers, and also because most DMs use free data to bootstrap their marketplace and attract the first ``buyers'' and then commercial sellers. 

The significant monthly growth rate we have seen at AWS and other DMs makes us believe that in the future the paid catalogue of most DMs is bound to grow and therefore, we will continue monitoring them to see how they evolve. We are also working on extending our feature importance analysis to become a price recommendation tool for new data products.

% The following two commands are all you need in the
% initial runs of your .tex file to
% produce the bibliography for the citations in your paper.
\bibliographystyle{abbrv}

% vldb_sample.bib is the name of the Bibliography in this case
% You must have a proper ".bib" file
%  and remember to run:
% latex bibtex latex latex
% to resolve all references

\vspace{-5mm}
\appendix{}
\section{Methodology} 

\subsection{Regression models}
\label{appendix:RegressionModels}
We have tested variations of 9 different regressors with different values for their main parameters (e.g., num. of estimators, depth, etc.) as included in the Scikit-learn~\cite{Pedregosa11} Python library, and inputs of different vocabulary lengths. Such models work with the log instead of the absolute value of product prices as the dependent variable so as to normalize the distribution of prices and avoid negative price predictions. We were hoping to find at least 3 models that produce sufficiently accurate price predictions, measured as the $R^2 score$ of their output with respect to actual prices.

To reduce the complexity of each  model, we removed low-value features, i.e., those that had a negative leave-one-out (LOO) value, provided the accuracy of the model was not negatively affected. A feature having negative LOO value means that the model improved its average accuracy in 10 random executions for different train and test data splits when such feature was removed from the input matrix. Finally, we performed a cross-validation to check the variance of the accuracy of the model when training and testing in 5-folds, and 20-random training-test splits of the input data.

We found that three target models worked reasonably well (i.e., they yield an $R^2$ score greater or equal to 0.70), namely random forest~\cite{Breiman01}, k-Nearest Neighbours~\cite{Kramer11}, and Gradient Goosting~\cite{Friedman00, Mason99} regression models. On the contrary, we discarded linear, Elastic-Net~\cite{Zou05}, Ridge~\cite{Hoerl00}, bayesian Ridge~\cite{MacKay92}, and Lasso~\cite{Tibshirani96} regressions even though they worked well in specific simulations. 

In addition, we also tested a Deep Neural Network regressor using the TensorFlow~\cite{tensorflow} and Keras~\cite{Keras} libraries. We followed all common good practices recommended for such activity by first standardizing the input data. We tested RELU/Leaky RELU activation functions for all hidden layers, and a Linear activation function for the output layer. As loss function we used the MAE. To avoid overfitting we randomly applied Drop-out between training epochs and to avoid dying/exploding neurons we also applied Batch normalization between all layers. We used the Adam optimizer~\cite{Kingma15} with a tuned learning rate decay to train the model faster at the beginning and then decrease the learning rate with further epochs to make training more precise. Finally, we used Callbacks to stop the training at the optimal epoch. 

\begin{table}[t]
\caption{Results of price prediction models}
%\vspace{-3mm}
\label{tab:ScoresRegressionModels}
\begin{small}
\resizebox{\columnwidth}{!}{%
\begin{tabular}{l||c|c|c||c|c|c||c|c|c||c|c|c}
\toprule
\hline
\multirow{2}{*}{\textbf{Model}} & \multicolumn{3}{c||}{\textbf{Financial}} & \multicolumn{3}{c||}{\textbf{Marketing}} & \multicolumn{3}{c||}{\textbf{Healthcare}} & \multicolumn{3}{c}{\textbf{All}} \\ \cline{2-13} 
    & \boldmath$R^2$ & \textbf{MAE}  & \textbf{MSE}  & \boldmath$R^2$ & \textbf{MAE}  & \textbf{MSE}  & \boldmath$R^2$ & \textbf{MAE}  & \textbf{MSE} & \boldmath$R^2$ & \textbf{MAE}  & \textbf{MSE} \\ \hline
\textbf{RF}&0.85&0.2&0.14&0.86&0.21&0.13&0.78&0.25&0.15&0.84&0.23&0.16\\ \hline
\textbf{kN}&0.78&0.31&0.26&0.74&0.33&0.24&0.77&0.26&0.17&0.69&0.37&0.31\\ \hline
\textbf{GB}&0.82&0.23&0.16&0.8&0.28&0.19&0.73&0.27&0.19&0.79&0.3&0.22\\ \hline
\textbf{DNN}               & 0.73                    & 0.33    & 0.35    & 0.77                 & 0.30 & 0.22 & 0.68                    & 0.26    & 0.18 &  &  &  \\ \hline
\bottomrule
\end{tabular}
}
\end{small}
\end{table}

Table~\ref{tab:ScoresRegressionModels} presents a summary of the accuracy obtained by the different regressors by category of data products, including the $R^2$ score, the MAE and the mean squared error (MSE) with regards to the actual log prices.

For the sake of robustness, such results were consistent across subsequent 5-fold and SV executions of the models: $R^2$ score showed a standard deviation below $4\%$ of the average in each round. Note that due to the total (low) number of observations that we have in our datasets, DNN models are not recommended, nevertheless, we wanted to explore them since we believe that they will further improve our results as soon as we manage to increase the overall size of our datasets. Consequently, we avoid using any DNN model in the feature importance analysis.
%that we explain in the next section. 
\vspace{-3mm}
\subsection{Feature importance analysis}
\label{appendix:FeatureImportanceAnalysis}

On the one hand, we evaluated the importance of individual features using the following methods: 
\textbf{(1)} measuring the accuracy lost by randomly shuffling the values of a certain feature among samples (permutation importance~\cite{Strobl08}), and
\textbf{(2)} measuring the prediction accuracy lost when one individual feature is removed from the inputs (leave-one-out or LOO value)

We cross-validated our results in 5-fold executions of both methods and took averages in order to disregard features that showed to be important only in specific tests. As regards robustness, we compared the top-20 ranking of every individual test to the top-20 average ranking of that algorithm and category. It turns out that both rankings have at least 5 features in common in 95\% of the cases, and a median of 13 common individual features. 

On the other hand, we measured the importance of groups of features by using the following methods:\\
\noindent\textbf{(1)} Measuring the prediction accuracy lost when a group of features is removed from the input dataset (leave-one-out)\\
\noindent\textbf{(2)} Measuring the average (in 20 random train/test split executions) Shapley value of each group of features~\cite{Shapley52, Ghorbani19}

The Shapley value is defined as the average $R^2$ score added by combining the information of a certain group of features with every possible mix of the rest of groups. This is a well-known and widely-used concept in game theory, economics and ML~\cite{Ghorbani19}, and it is considered a `fair' method to distribute the gains obtained by cooperation. In our case, we applied the Shapley value to distribute the gains in accuracy of our regression models among the groups of features that contributed to achieving such an accuracy. Furthermore, we ran 5-fold feature importance analysis in the case of LOO, in a similar way as we did for individual features, and 20 calculations of the Shapley values for random 80/20 train/test splits of our input data. 

Table ~\ref{tab:RankingFeatureGroupsLOO} and table ~\ref{tab:ShapleyValueFeatureGroups} show the average LOO and Shapley values of such executions respectively. The standard deviation of Shapley values across executions is acceptable (average below 0.029 for financial and marketing datasets, 0.057 for healthcare-related data, and below 0.017 for all the data), and the ranking of relevant feature groups remains stable.

\end{document}